\documentclass{ieeetmlcn}
\usepackage{cite}
\usepackage{amsmath,amssymb,amsfonts}
\usepackage{algorithmic}
\usepackage{graphicx,color}
\usepackage{textcomp}
\usepackage{xcolor}
\usepackage{hyperref}

\usepackage{booktabs}
\usepackage{multirow}
\usepackage{makecell}
\usepackage{tabularx}
\usepackage{adjustbox}

\usepackage{bm}
\usepackage{mathtools}

\DeclareMathOperator{\CE}{CE}
\DeclareMathOperator{\TopK}{TopK}

\DeclareMathOperator{\GN}{GN}
\DeclareMathOperator{\SiLU}{SiLU}
\DeclareMathOperator{\Conv}{Conv}

\DeclareMathOperator{\sgn}{sgn}

\usepackage[caption=false,font=footnotesize]{subfig}

\hypersetup{hidelinks=true}
\usepackage{algorithm,algorithmic}
\def\BibTeX{{\rm B\kern-.05em{\sc i\kern-.025em b}\kern-.08em
    T\kern-.1667em\lower.7ex\hbox{E}\kern-.125emX}}
\AtBeginDocument{\definecolor{tmlcncolor}{cmyk}{0.93,0.59,0.15,0.02}\definecolor{NavyBlue}{RGB}{0,86,125}}

\def\OJlogo{\vspace{-4pt}\includegraphics[height=18pt]{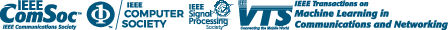}}
\def\seclogo{\vspace{10pt}\includegraphics[height=18pt]{new_logo}}

\def\authorrefmark#1{\ensuremath{^{\textbf{#1}}}}

\begin{document}
\receiveddate{XX Month, XXXX}
\reviseddate{XX Month, XXXX}
\accepteddate{XX Month, XXXX}
\publisheddate{XX Month, XXXX}
\currentdate{XX Month, XXXX}
\doiinfo{TMLCN.2022.1234567}

\markboth{Task-Oriented Candidate-Latent
Feedback for Coarse-to-Fine Sensing
in Distributed OFDM-ISAC Networks}{Shiv Shankar \textit{et al.}}

\title{Task-Oriented Candidate-Latent Feedback for Coarse-to-Fine Sensing in Distributed OFDM-ISAC Networks}

\author{Shiv Shankar \authorrefmark{1,2} Radha Krishna Ganti \authorrefmark{1}, J Klutto Milleth \authorrefmark{2}}
\affil{Indian Institute of Technology (IIT), Madras, Chennai India}
\affil{Centre of Excellence
in Wireless Technology (CEWiT), 
Chennai India}
\corresp{Corresponding author: Shiv Shankar (e-mail: shivshankar@cewit.org.in).}
\authornote{This work was funded in part by the Ministry of Electronics and Information Technology (MeitY), Government of India.}

\begin{abstract}
Future integrated sensing and communication (ISAC) architectures separate the sensing entity (SE) that acquires measurements from the sensing function (SF) that performs inference, creating a need for compact, task-oriented feedback on the SE--SF interface. Forwarding the raw channel frequency response (CFR) or full per-link delay--Doppler--azimuth--elevation (DDAE) tensor is prohibitively expensive, while peak-only reporting discards target-discriminative structure under clutter. We propose a learning-based coarse-to-fine sensing pipeline with candidate-latent feedback for single-target estimation. At the SE, a lightweight convolutional scorer produces a dense delay--Doppler proposal map from pilot-based OFDM channel estimates, and a learned encoder constructs $K$ compact $C$-dimensional candidate tokens by fusing per-candidate azimuth--elevation patches, normalized position, and confidence cues. The latents are uniformly quantized post-training to $b$ bits and transmitted under a finite budget $B_{\mathrm{fb}} = bKC + 18K + 16$ bits to the SF, which performs cross-candidate refinement, reranking, and joint four-parameter estimation. We characterize the $(K, C, b)$ design space and identify three operating points spanning payload range. On a ray-traced urban scene with static and dynamic clutter, these points achieve $96.33$--$98.88\%$ detection at $107$--$806$ bytes per coherent processing interval, corresponding to compression ratios of $1.2$--$9.2 \times 10^{4}$ over the $8$-bit DDAE magnitude tensor and reducing the SE--SF interface from multi-Gbit/s to sub-Mbit/s rates. Cross-scene evaluation on an independent campus-scale environment achieves $98.79$--$99.50\%$ detection and at-or-better angular accuracy without retraining, indicating that the learned representation captures target-relevant structure that transports across scenes of comparable or lower clutter density. Extension to the multi-target regime, quantization-aware training, and full multistatic fusion are identified as natural next steps.
\end{abstract}


\maketitle

\section{INTRODUCTION}
\label{sec:introduction}

\IEEEPARstart{I}{ntegrated} sensing and communication (ISAC) has emerged as a foundational capability for beyond-5G and sixth-generation (6G) wireless networks, enabling communication infrastructure to simultaneously support high-rate data delivery and fine-grained environment awareness by sharing spectrum, hardware, and signaling resources~\cite{Liu2022DualFunctional,Zhang2021OverviewISAC,Lu2024RecentAdvancesISAC}. By co-designing the sensing and communication functionalities rather than treating them as isolated subsystems, ISAC offers substantial gains in spectral efficiency, hardware economy, and operational synergy~\cite{Cui2021IntegratingSensing,Liu2020JointRadarComm}. These advantages have propelled ISAC to the forefront of 6G research agendas, with the International Telecommunication Union (ITU) identifying it as one of the six core usage scenarios for International Mobile Telecommunications~2030 (IMT-2030)~\cite{Liu2022DualFunctional,Wymeersch2022Integration}. The growing consensus is that sensing should no longer be viewed as an add-on to communication but rather as a \emph{native} function of future wireless systems, tightly integrated into the network architecture from the outset~\cite{Zhang2021OverviewISAC,Strinati2025DISAC}.

This architectural shift is particularly consequential for \emph{networked sensing}, in which the entity that acquires sensing measurements and the function that performs the final inference need not be collocated. Dense cellular deployments, distributed radio nodes, and communication-native pilot structures can be leveraged not only for connectivity, but also for localization, tracking, imaging, and scene understanding across a wide area~\cite{Cui2024NetworkLevelPerspective,Li2024NetworkedISAC,Strinati2025DISAC}. In such settings, different base stations (gNBs), user equipment (UEs), or specialized sensing nodes may observe complementary views of the same target or environment, and the resulting measurements must be compressed and transported through the network before a final inference can be made. This inherently distributed topology introduces a new design dimension that goes beyond traditional single-node radar signal processing: in addition to the estimation algorithm itself, the system must provide principled interfaces for transferring sensing information across the network under communication constraints~\cite{Cui2024NetworkLevelPerspective,Strinati2025DISAC,Wymeersch2022Integration}.

Recent standardization activity further reinforces this distributed perspective. Within the Third Generation Partnership Project (3GPP), Release~19 introduced both the stage-1 study item TR~22.837~\cite{3GPPTR22837} and the stage-1 specification TS~22.137~\cite{3GPPTS22137} for integrated sensing and communication, defining the service-level requirements and reference scenarios for cellular ISAC. The ongoing Release~20 stage-2 architectural study TR~23.700-14~\cite{3GPPTR2370014} is investigating function decomposition, service exposure, and end-to-end operational procedures, explicitly distinguishing between a \emph{Sensing Entity}~(SE) that acquires and preprocesses measurements and a \emph{Sensing Function}~(SF) that executes the final sensing algorithm. In parallel, recent IEEE architectural discussions~\cite{Strinati2025DISAC,Cui2024NetworkLevelPerspective} have argued that sensing should be treated as a networked function where radio nodes observe, preprocess, and exchange sensing information before inference is performed. This makes the ISAC design problem inherently \emph{two-sided}: one side acquires and compresses the measurement, while the other side consumes a transported representation to execute the downstream task. The central unresolved question is therefore not only \emph{how to sense}, but also \emph{what intermediate representation} should be transmitted across the network to enable accurate inference at the receiver under a finite feedback budget.

At the same time, practical ISAC sensing is fundamentally more difficult than the idealized settings often assumed in early studies. Real propagation environments contain strong static reflections from buildings and infrastructure, dynamic clutter from moving vehicles and pedestrians, and site-dependent multipath, all of which can overlap with the target response in delay, Doppler, and angle~\cite{Wei2025ChannelModelSurvey,Luo2024ClutterEnvironment,Yang2024ISACChannelMeas}. Recent channel-modeling and clutter-aware sensing studies have therefore argued that realistic ISAC pipelines must explicitly model the target--clutter superposition rather than rely on idealized sparse-target assumptions~\cite{Wei2025ChannelModelSurvey,Zhang2022GeneralChannelModel}, particularly in communication-centric sensing tied to pilot-based OFDM/NR channel measurements~\cite{Gao2022CFR5GNR,Sturm2011WaveformDesign}. In such settings, forwarding the raw per-link measurement to a central node is prohibitively expensive in bandwidth, while reducing it to a few hard spectral peaks discards target-discriminative structure needed for robust inference. This tension between feedback cost and information preservation motivates the search for compact, task-oriented sensing representations that retain target-specific structure under a finite feedback budget.

\subsection{Related Work}

% \subsubsection{Model-Based ISAC Parameter Estimation}
A substantial body of work has established that communication-compatible OFDM and multiple-input multiple-output (MIMO) waveforms can support accurate joint estimation of target parameters without requiring a dedicated radar waveform~\cite{Sturm2011WaveformDesign,Liu2018DFRC_MIMO}. The dual-functional radar-communication (DFRC) paradigm has shown that carefully designed ISAC transmit signals can achieve near-optimal radar estimation performance while simultaneously maintaining high-quality data delivery~\cite{Liu2018DFRC_MIMO,Liu2020JointRadarComm}. Representative recent advances include joint angle--range--velocity estimation methods for MIMO-OFDM ISAC systems that exploit the full coherent processing interval and achieve significantly improved root-mean-squared error (RMSE) by jointly processing the received echoes across subcarriers, symbols, and antenna elements~\cite{Xiao2024JointARV}. Radar-assisted predictive beamforming approaches further demonstrate that sensing-derived state information can be fed back into the communication subsystem to enable proactive resource allocation for vehicular and high-mobility scenarios~\cite{Liu2020RadarAssistedBeamforming,Yuan2021SpatioTemporalDualFunction}.

Importantly, practical clutter-aware ISAC sensing frameworks explicitly model the received channel as the superposition of dynamic target returns and environmental clutter~\cite{Luo2024ClutterEnvironment}. These methods typically rely on explicit filtering or subspace/spectrum-based processing---for example, mean-phasor cancellation for removing quasi-static environmental clutter followed by joint detection and angle/range/velocity estimation. More broadly, recent channel modeling studies have emphasized that realistic ISAC sensing must account not only for the target path, but also for static and dynamic background components, clutter radar cross section (RCS), and site-specific multipath, especially in communication-native deployments that inherit the structure of OFDM pilots and 3GPP-style channels~\cite{Wei2025ChannelModelSurvey,Zhang2022GeneralChannelModel}. Taken together, this line of work has established that realistic ISAC target estimation requires explicit attention to clutter, background structure, and communication-compatible signal models rather than relying on idealized sparse-target assumptions alone.

% \subsubsection{Learning-Based ISAC Receivers}
In parallel, learning-based ISAC methods have begun to move beyond purely model-driven sensing pipelines by jointly learning signal detection and target estimation from communication-derived measurements~\cite{Hu2024TwoStageISACReceiver,Jiang2024ISACNET}. The two-stage SIMO-OFDM ISAC receiver of Hu~\emph{et al.}~\cite{Hu2024TwoStageISACReceiver} combines learning-based symbol detection with subsequent target-parameter estimation, demonstrating that end-to-end training can mitigate error propagation across the detection and sensing stages. ISAC-NET~\cite{Jiang2024ISACNET} employs a model-driven deep unfolding architecture to jointly perform passive sensing and communication demodulation, reporting significant sensing gains over conventional two-dimensional DFT-based processing. More broadly, the model-based deep learning paradigm has shown that physics-informed neural network architectures can achieve strong performance in wireless estimation tasks while retaining interpretability and convergence guarantees~\cite{Shlezinger2023ModelBased,OShea2017IntroductionDLPHY}. These results collectively indicate that learning can improve robustness to demodulation errors and can exploit structured communication measurements more effectively than rigid hand-crafted pipelines.

However, existing learning-based ISAC receivers are still largely \emph{centralized}: the learned model is typically assumed to operate where the measurement is available, and the main design question is how to estimate the target parameters accurately from that measurement. They do not directly address the increasingly important distributed setting in which the sensing entity and the final inference function are separated and only a limited amount of sensing information can be conveyed across the network. In particular, prior ISAC estimation works do not yet provide a learned, task-oriented feedback representation that is rich enough to retain target-discriminative structure under clutter, yet compact enough to replace raw DDAE tensor or channel frequency response (CFR) forwarding. This missing representation layer is precisely the gap addressed in the present work.

% \subsubsection{Deep Learning for CSI Feedback}
A closely related design philosophy has already matured in massive-MIMO channel state information (CSI) feedback. CsiNet~\cite{Wen2018CsiNet} demonstrated that an autoencoder-based encoder--decoder pair compresses downlink CSI into a small set of learned latent space at lower overhead and higher reconstruction quality than classical compressive sensing. CsiNet-LSTM~\cite{Wang2019CSINetLSTM} extends this to time-varying channels, DeepCMC~\cite{Mashhadi2021DeepCMC} introduces fully convolutional distributed compression with variable-rate feedback and channel-adaptive quantization, and more recent designs~\cite{Liang2024LowRateCSIFeedback,Guo2022DLOverview} learn the feedback representation jointly with the end task rather than via a fixed hand-crafted transform. The unifying insight is that \emph{what} is fed back matters as much as \emph{how much}: a learned latent that discards irrelevant channel structure while preserving task-critical features dramatically outperforms classical feedback at the same bit budget.

% \subsubsection{Semantic and Task-Oriented Communication}
A parallel paradigm shift is underway in semantic and task-oriented communications~\cite{Yang2023SemComFutureInternet,Lu2024SemanticsEmpowered,Chaccour2025LessDataMoreKnowledge}. Moving beyond Shannon-theoretic bit-level reconstruction, this framework advocates transmitting representations that preserve the task utility of the source under bandwidth, delay, and robustness constraints~\cite{Lu2023RethinkingSemantic,Guo2025SemComNetworksSurvey}. Task-oriented methods learn compact latent features optimized for a downstream objective (classification, decision, control, or retrieval) while discarding source details irrelevant to that objective~\cite{Xie2022TaskOrientedMultiUser,Bourtsoulatze2019DeepJSCC,Jankowski2020WirelessImageTransmission}. Applied to distributed ISAC, this principle dictates that the SE should extract and transmit a latent representation capturing the target-relevant content of the observation rather than attempting to reproduce the full measurement at the receiver.

% \subsection{Motivation and Research Gap}

The preceding discussion identifies a clear gap: existing ISAC sensing methods---both model-based and learning-based---are designed for centralized operation and do not provide a learned feedback representation that can bridge the SE--SF split under a finite bit budget. At the same time, the CSI feedback and semantic communication literatures have demonstrated that learned, task-oriented latent representations can dramatically outperform hand-crafted feedback at the same overhead. Combining these insights, the goal of this work is to learn a feedback representation for ISAC sensing that is simultaneously (i)~\emph{compact} enough for limited-rate transport, (ii)~\emph{specific} enough to preserve target-related structure under clutter and background variation, and (iii)~\emph{directly consumable} by the downstream inference function without requiring reconstruction of the full measurement tensor. This representation-centric viewpoint, which is well-established in CSI compression~\cite{Wen2018CsiNet,Mashhadi2021DeepCMC,Guo2022DLOverview} and increasingly central in semantic communication~\cite{Yang2023SemComFutureInternet,Lu2024SemanticsEmpowered,Chaccour2025LessDataMoreKnowledge}, remains largely unexplored in ISAC target estimation.

Concretely, to the best of our knowledge, no prior ISAC estimator---model-based or learning-based---jointly estimates
delay, Doppler, azimuth, and elevation over a UPA in a bistatic
ray-traced scene with static and dynamic clutter under a finite
per-link feedback budget. Existing learning-based estimators address
subsets of this problem: \cite{Hu2024TwoStageISACReceiver} estimates
AoA and delay only; \cite{Jiang2024ISACNET} estimates range and
velocity with no angular output; \cite{Xiao2024JointARV,HuZelin2024}
estimate range/velocity/azimuth over ULAs without elevation; and
the closest model-based work \cite{Luo2024ClutterEnvironment}
handles angle/distance/velocity but assumes explicit static-clutter
filtering and does not address feedback compression. More
fundamentally, none of the above works adopts a decentralized
SE--SF architecture in which a target-specific learned latent
representation is transmitted from the sensing entity to the
sensing function; the present work targets exactly this regime.

\subsection{Contributions}

This paper proposes task-oriented candidate-latent feedback for coarse-to-fine sensing in distributed OFDM-ISAC networks. In the considered SE--SF architecture, the sensing entity (SE) avoids forwarding the full delay--Doppler--azimuth--elevation (DDAE) tensor to the sensing function (SF). Instead, it generates a compact set of delay--Doppler candidates and encodes each candidate into a learned latent token that preserves task-relevant sensing information under a finite feedback budget. The SF then uses only the received candidate latents and indices to perform reranking and refined target-parameter estimation. The main contributions are as follows.

\begin{itemize}
    \item \textit{Task-oriented candidate-latent feedback interface:}
    We formulate a compact SE--SF feedback interface in which each delay--Doppler candidate is represented by a learned latent token. Unlike raw DDAE forwarding or peak-only reporting, the proposed representation preserves target-discriminative angular and delay--Doppler structure while substantially reducing the feedback payload.

    \item \textit{Finite-rate feedback model and operating-point design:}
    We introduce a quantized candidate-latent feedback model with bit-exact payload accounting,
    \(B_{\mathrm{fb}}=bKC+18K+B_{\mathrm{side}}\), where \(K\) is the number of candidates, \(C\) is the latent dimension, and \(b\) is the number of quantization bits per latent component. This enables direct payload--accuracy tradeoff analysis and systematic selection of compact operating points.

    \item \textit{Coarse-to-fine sensing architecture for distributed OFDM-ISAC:}
    We develop an end-to-end estimator that first generates high-recall delay--Doppler proposals at the SE, constructs candidate tokens from angular patches, candidate coordinates, and confidence cues, and then performs set-based refinement, reranking, and four-parameter estimation at the SF. The design explicitly separates local measurement processing at the SE from downstream inference at the SF.
    
    \item \textit{Ray-traced validation under clutter and cross-scene testing:}
    We evaluate the proposed framework on a realistic ray-traced urban ISAC dataset with static and dynamic clutter, benchmark it against classical peak-selection baselines, and study the effects of candidate count, latent dimension, quantization width, and architectural components. The selected operating points achieve high detection accuracy at payloads ranging from 107 to 806 bytes per sample, tolerate 6-bit post-training quantization with negligible degradation, and transfer without retraining to an unseen campus-scale scene.
\end{itemize}

\subsection{Paper Organization}
The rest of the paper is organized as follows. Section~\ref{sec:system_model} presents the system model and sensing signal representation. Section~\ref{sec:method} develops the proposed coarse-to-fine estimator and candidate-latent feedback interface. Section~\ref{sec:setup} presents the simulation setup. Section~\ref{sec:results} reports the main results, including the comparison with classical peak-selection baselines, the $(K,C)$ operating-point sweep at FP32, post-training quantization sensitivity, cross-scene validation on Scene-IITM, and component ablations. Section~\ref{sec:conclusion} concludes the paper and outlines future work.

\begin{figure}[!t]
  \centering
  \includegraphics[width=1\linewidth, trim={110 210 90 90}, clip]
  {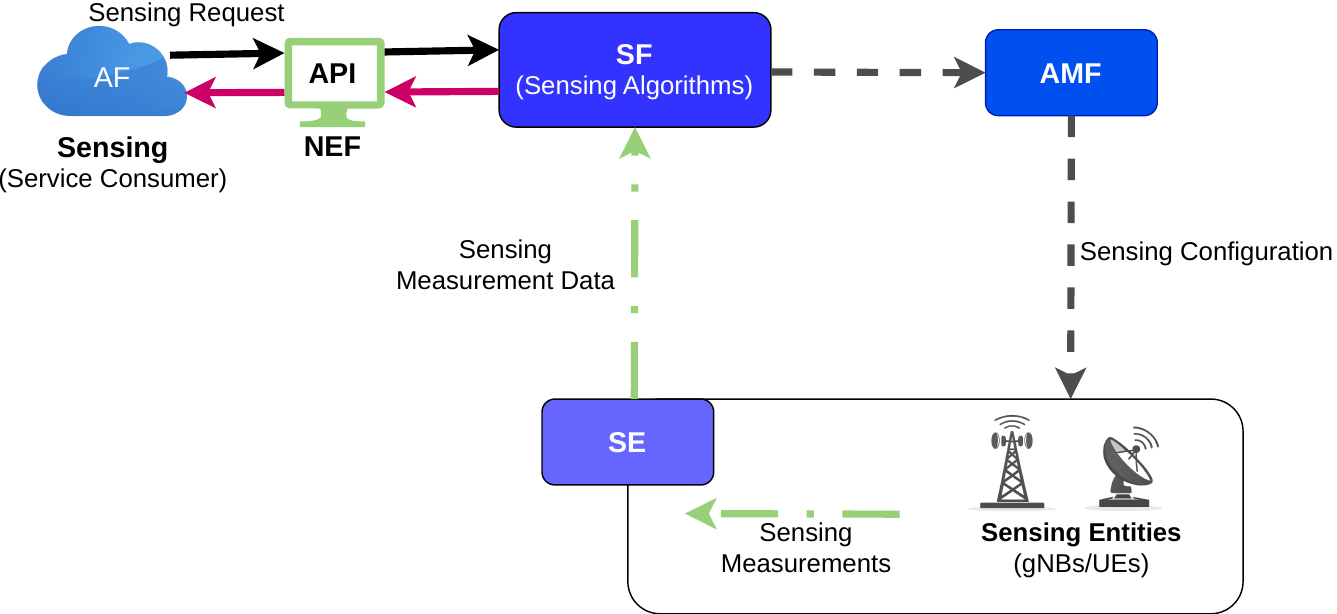}
\caption{High-level system architecture for distributed ISAC sensing, showing the SE--SF split with sensing measurement transport via the network exposure function (NEF).}

\label{fig:ISAC_Architecture_drawio}
\end{figure} 

\section{SYSTEM MODEL}
\label{sec:sysmodel}
\begin{figure}[!t]
  \centering
\includegraphics[width=0.95\linewidth, trim={0 0 0 0}, clip]{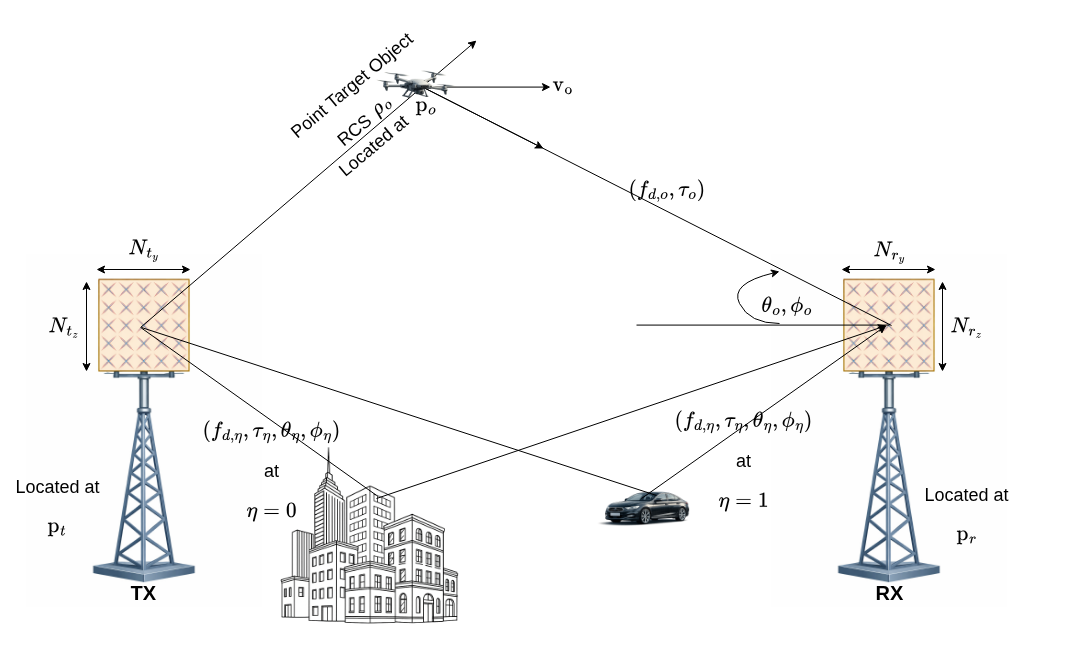}
\caption{Bistatic sensing geometry with TX and RX UPAs, a point target at $\mathbf{p}_o$, and the associated delay, Doppler, AoD, and AoA parameters.}
\label{fig:system_model}
\end{figure}

\label{sec:system_model}

\subsection{Geometry, Arrays, and Far-Field Assumption}
Consider a multistatic (bistatic) ISAC network (Fig.~\ref{fig:system_model}) with transmitter (TX) at positions $\mathbf{p}_t$, receiver (RX) at positions $\mathbf{p}_r$, and point targets at $\mathbf{p}_o$ with radar cross section (RCS) $\rho_o$.
The model allows monostatic operation as the special case $\mathbf{p}_t=\mathbf{p}_r$, but the numerical experiments focus on distributed (bistatic) deployments where TX and RX are physically separated.

Each TX employs an $N_{t_y}\!\times N_{t_z}$ uniform planar array (UPA) lying in the $yz$-plane, with inter-element spacing $d_{ant}=\lambda/2$, where $\lambda$ is the carrier wavelength. Let $N_t=N_{t_y}N_{t_z}$ denote the number of TX elements. Similarly, each RX employs an $N_{r_y}\!\times N_{r_z}$ UPA (also in its local $yz$-plane) with $N_r=N_{r_y}N_{r_z}$ elements. Each RX defines a local coordinate system (LCS) where the $x$-axis is normal (broadside) to the UPA, the $z$-axis aligns with the array’s top row, and the array lies in the $yz$-plane.

Targets are placed uniformly at random within a bounded 3-D region
$x_{\min}<x<x_{\max}$, $y_{\min}<y<y_{\max}$, and $z_{\min}<z<z_{\max}$.
For each TX--target--RX propagation path, we assume a \emph{practical far-field} condition: while the distance may not strictly exceed the Fraunhofer distance (on the order of $2D_{\max}^2/\lambda$ with $D_{\max}=\max\{D_t,D_r\}$ the larger aperture), planar-wavefront and uniform spherical-spreading approximations remain accurate. Hence, each path is modeled as a plane wave whose AoD/AoA is constant across the corresponding UPA.

All TXs and RXs are assumed perfectly synchronized in time and frequency.

\subsection{Target Motion and RCS}
Target velocities are drawn uniformly in $\mathbb{R}^3$ subject to $\|\mathbf{v}_o\|<v_{\max}$.
The target RCS (in dBsm) is modeled as
$\rho_{o,\mathrm{dBsm}}\sim \mathcal{N}(\rho_m,\sigma_\rho^2)$,
and $\rho_o = 10^{\rho_{o,\mathrm{dBsm}}/10}$.

\subsection{Clustered Multipath Channel Model (TR~38.901)}
We adopt the clustered delay-line structure of 3GPP TR~38.901~\cite{3gpp-tr-38901}. For a given TX/RX antenna pair $(n_{\mathrm{tx}},n_{\mathrm{rx}})$, the continuous-time channel impulse response (CIR) is
\begin{equation}
\begin{aligned}
h_{n_{\mathrm{tx}},n_{\mathrm{rx}}}(\tau,t)
&=\sum_{\eta=1}^{N_{\mathrm{path}}} a_\eta\,
e^{j\frac{2\pi}{\lambda}\hat{\mathbf{r}}_{\mathrm{rx},\eta}^{\mathsf T}\mathbf{d}_{n_{\mathrm{rx}}}}\,
e^{j\frac{2\pi}{\lambda}\hat{\mathbf{r}}_{\mathrm{tx},\eta}^{\mathsf T}\mathbf{d}_{n_{\mathrm{tx}}}}\,\\
&\quad\times e^{j2\pi f_{d,\eta}t}\,
\delta\!\bigl(\tau-\tau_{n_{\mathrm{tx}},n_{\mathrm{rx}},\eta}\bigr),
\end{aligned}
\label{eq:cir_38901}
\end{equation}
where $N_{\mathrm{path}}$ is the number of paths, $a_\eta$ is the complex path gain, $\hat{\mathbf{r}}_{\mathrm{tx},\eta}$ and $\hat{\mathbf{r}}_{\mathrm{rx},\eta}$ are unit departure/arrival direction vectors (in the TX/RX LCS, respectively), $\mathbf{d}_{n_{\mathrm{tx}}}$ and $\mathbf{d}_{n_{\mathrm{rx}}}$ are the element position vectors, and $\tau_{n_{\mathrm{tx}},n_{\mathrm{rx}},\eta}$ and $f_{d,\eta}$ are the path delay and Doppler shift.

\subsection{ISAC Channel Decomposition: Target + Background}
We decompose the overall ISAC channel into a target-induced component and an environmental (background) component:
\begin{equation}
h(\tau,t)=h_o(\tau,t)+h_{\mathrm{bg}}(\tau,t).
\label{eq:isac_decomp}
\end{equation}
For a point target, the target component is constructed by convoluting TX$\to$target and target$\to$RX responses and scaling by the target RCS:
\begin{equation}
h_o(\tau,t)=\sqrt{\rho_o}\; h_{to}(\tau,t) * h_{or}(\tau,t).
\label{eq:target_conv}
\end{equation}
In simulations, we use NVIDIA Sionna to generate CIRs via 3-D ray tracing for $h_{\mathrm{bg}}(\tau,t)$, producing site-specific path parameters while preserving the cluster/ray structure of~\eqref{eq:cir_38901}.

\subsection{OFDM Representation}
We consider an orthogonal frequency-division multiplexing (OFDM) system with $N_{sc}$ subcarriers, subcarrier spacing $\Delta f$, and $L$ OFDM symbols.
Let $T_{\mathrm{sym}}=1/\Delta f$ and $t_l=lT_{\mathrm{sym}}$.
The baseband channel frequency response (CFR) is
\begin{equation}
\begin{aligned}
h[n_{sc},l]&=\int_{-\infty}^{\infty} h(\tau,t_l)\, e^{-j2\pi n_{sc}\Delta f\tau}\, d\tau,\\
&\quad n_{sc}=0,\dots,N_{sc}{-}1,\; l=0,\dots,L{-}1.
\label{eq:cfr_def}
\end{aligned}
\end{equation}
The target contribution in frequency can be expressed as
\begin{equation}
h_o[n_{sc},l]=\sqrt{\rho_o}\; h_{to}[n_{sc},l]\; h_{or}[n_{sc},l].
\label{eq:cfr_target_prod}
\end{equation}

\subsection{Array Steering Vectors and Beamforming}
For a UPA in the $yz$-plane with spacing $d_{ant}=\lambda/2$, index elements by $(n_y,n_z)$ with
$n_y\in\{0,\dots,N_y{-}1\}$ and $n_z\in\{0,\dots,N_z{-}1\}$, and position vector
$\mathbf{d}_{ant,n_y,n_z}=[0,\; n_y d_{ant},\; n_z d_{ant}]^{\mathsf T}$ in the array LCS.

We parameterize a plane-wave direction by elevation--azimuth $(\theta,\phi)$ such that its direction cosines along $y$ and $z$ in the array LCS are
$u_y(\theta,\phi)=\cos\theta\,\sin\phi$ and $u_z(\theta,\phi)=\sin\theta$.
Then the (TX or RX) steering vector $\bigl[\mathbf{a}(\theta,\phi)\bigr]_{n_y,n_z}$ can be written compactly as
\begin{equation}
\exp\!\left(j\frac{2\pi}{\lambda}\Bigl(u_y(\theta,\phi)\,n_y d_{ant} + u_z(\theta,\phi)\,n_z d_{ant}\Bigr)\right).
\label{eq:steering_vec}
\end{equation}

TX beamforming uses a single spatial layer $x[n_{sc},l]\in\mathbb{C}^{1}$ shared across all TX elements with a (static or slowly updated) weight vector
$\mathbf{w}_t\in\mathbb{C}^{N_t}$ normalized as $\|\mathbf{w}_t\|_2^2=N_t$:
\begin{equation}
\tilde{\mathbf{x}}[n_{sc},l]=\mathbf{w}_t\,x[n_{sc},l].
\label{eq:tx_bf}
\end{equation}
The effective TX array factor toward direction $(\psi,\xi)$ is
\begin{equation}
\gamma(\psi,\xi)=\mathbf{a}_t^{T}(\psi,\xi)\,\mathbf{w}_t.
\label{eq:tx_array_factor}
\end{equation}

\subsection{Target Path: Bistatic Delay and Doppler}
For a bistatic TX--target--RX geometry, the target round-trip delay is
\begin{equation}
\tau_o=\frac{\|\mathbf{p}_o-\mathbf{p}_t\|+\|\mathbf{p}_r-\mathbf{p}_o\|}{c},
\label{eq:bistatic_delay}
\end{equation}
and the bistatic Doppler shift is
\begin{equation}
\begin{aligned}
&f_{d}=\frac{\mathbf{v}_o^{\mathsf T}\hat{\mathbf{r}}_{to}-\mathbf{v}_o^{\mathsf T}\hat{\mathbf{r}}_{or}}{\lambda},\\
&
\hat{\mathbf{r}}_{to}=\frac{\mathbf{p}_o-\mathbf{p}_t}{\|\mathbf{p}_o-\mathbf{p}_t\|},
\quad
\hat{\mathbf{r}}_{or}=\frac{\mathbf{p}_r-\mathbf{p}_o}{\|\mathbf{p}_r-\mathbf{p}_o\|}.
\label{eq:bistatic_doppler}
\end{aligned}
\end{equation}

Let $(\psi_o,\xi_o)$ denote the AoD from the TX toward the target and $(\theta_o,\phi_o)$ the AoA at the RX from the target (both in the corresponding LCSs).
With overall target path gain $a_o$ and TX factor $\gamma_o=\gamma(\psi_o,\xi_o)$, the target CFR at RX element $(n_{r_y},n_{r_z})$ can be written as
\begin{equation}
\begin{aligned}
H_o(n_{r_y},n_{r_z},n_{sc},l)
=\sqrt{\rho_o}\;\gamma_o\,a_o\;
e^{-j2\pi n_{sc}\Delta f\,\tau_o}\;
e^{j2\pi f_{d}lT_{\mathrm{sym}}}\;\\
\quad\times\exp\!\left(j\frac{2\pi}{\lambda}\Bigl(u_y(\theta_o,\phi_o)\,n_{r_y}d_{ant} + u_z(\theta_o,\phi_o)\,n_{r_z}d_{ant}\Bigr)\right).
\end{aligned}
\label{eq:target_cfr}
\end{equation}

\paragraph{Monostatic Special Case.}
When $\mathbf{p}_t=\mathbf{p}_r$, the range, delay, and Doppler reduce to
\begin{equation}
\begin{aligned}
&R=\|\mathbf{p}_o-\mathbf{p}_r\|,
\qquad
\tau=\frac{2R}{c},\\
&f_{d}=\frac{2\,\mathbf{v}_o^{\mathsf T}\hat{\mathbf{r}}}{\lambda},
\quad
\hat{\mathbf{r}}=\frac{\mathbf{p}_o-\mathbf{p}_r}{R}.
\label{eq:mono}
\end{aligned}
\end{equation}

\subsection{Background Channel}
The background component aggregates all reflections not deliberately produced by the target (e.g., buildings and ground). It is treated as structured interference during sensing and as a quasi-static channel for communication. Let there be $N_{b\mathrm{path}}$ background paths. For the $\eta$-th background path with AoD $(\psi_\eta,\xi_\eta)$ and AoA $(\theta_\eta,\phi_\eta)$, delay $\tau_\eta$, Doppler $f_{d,\eta}$, and gain $a_\eta$, the CFR at RX element $(n_{r_y},n_{r_z})$ is
\begin{equation}
\begin{aligned}
&H_{\mathrm{bg},\eta}(n_{r_y},n_{r_z},n_{sc},l)
=a_\eta\,\gamma(\psi_\eta,\xi_\eta)\;\\
&\quad\times\exp\!\left(j\frac{2\pi}{\lambda}\Bigl(u_y(\theta_\eta,\phi_\eta)\,n_{r_y}d_{ant} + u_z(\theta_\eta,\phi_\eta)\,n_{r_z}d_{ant}\Bigr)\right) \\
&\quad\times e^{-j2\pi n_{sc}\Delta f\,\tau_\eta}\;
e^{j2\pi f_{d,\eta}lT_{\mathrm{sym}}}.
\end{aligned}
\label{eq:bg_cfr}
\end{equation}
The total background CFR is
\begin{equation}
H_{\mathrm{bg}}(n_{r_y},n_{r_z},n_{sc},l)=\sum_{\eta=1}^{N_{b\mathrm{path}}} H_{\mathrm{bg},\eta}(n_{r_y},n_{r_z},n_{sc},l).
\label{eq:bg_sum}
\end{equation}
The background includes both near-zero Doppler (static) and nonzero Doppler (moving clutter) components and occupies a single PRS port in the communication system.

\subsection{Channel Tensor, Received Signal, and Pilot-Based LS Estimate}
For a given TX--RX pair, stack across the RX array and OFDM time--frequency grid to form the channel tensor
$\mathbf{H}\in\mathbb{C}^{N_{r_y}\times N_{r_z}\times N_{sc}\times L}$:
\begin{equation}
\mathbf{H}=\mathbf{H}_o+\mathbf{H}_{\mathrm{bg}}.
\label{eq:H_total}
\end{equation}
The received signal tensor is
\begin{equation}
\mathbf{Y}=\mathbf{H}\odot\mathbf{X}+\mathbf{N},
\label{eq:rx_sig}
\end{equation}
where $\odot$ denotes elementwise multiplication over $(n_{sc},l)$, $\mathbf{X}$ is the transmitted resource grid (shared across RX antennas), and $\mathbf{N}$ is additive noise.

\begin{figure}[!t]
  \centering
  \includegraphics[width=0.8\linewidth, trim={0 0 0 0}, clip]{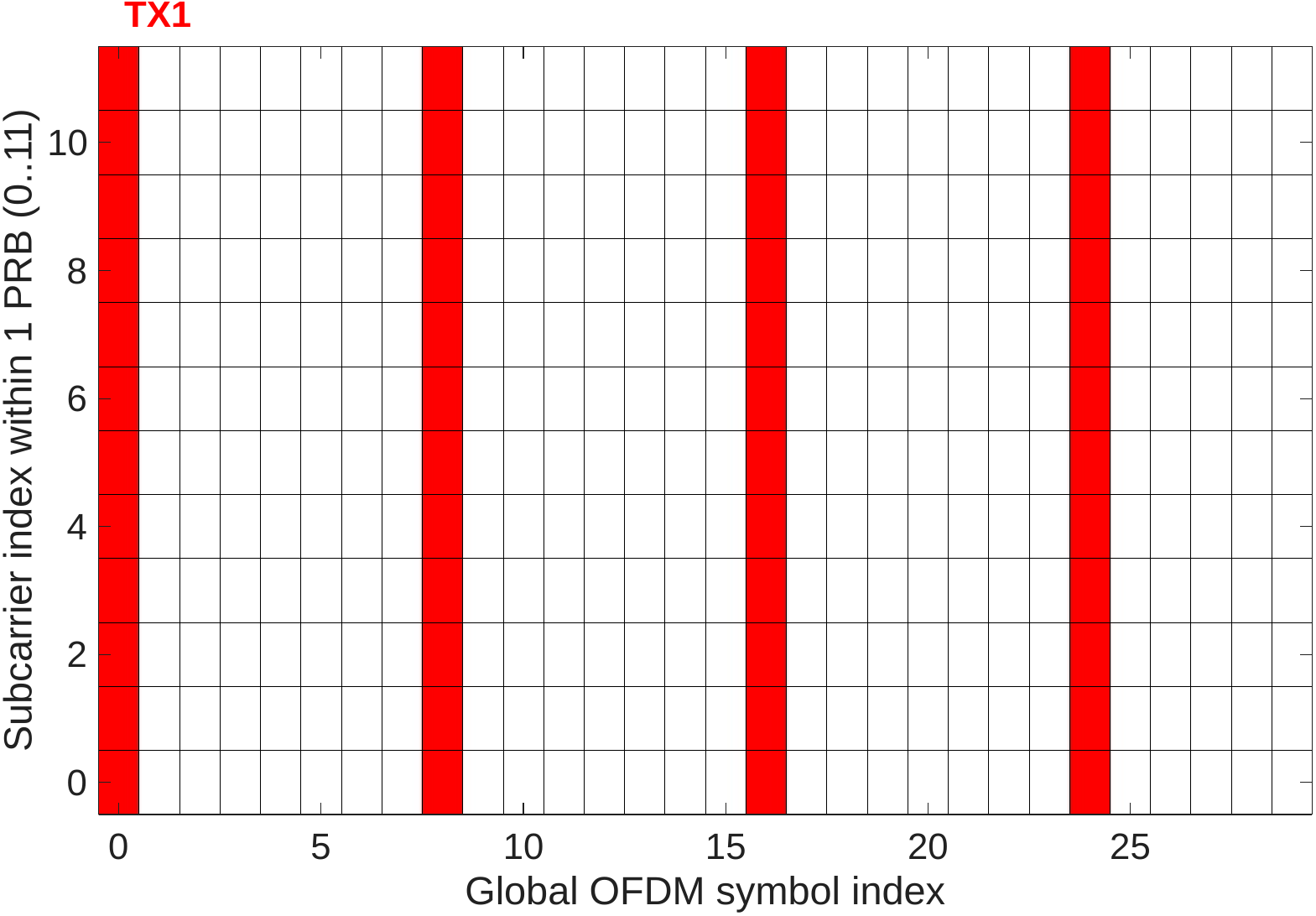}
  \caption{Positioning-reference-signal (PRS) resource allocation: one PRS OFDM symbol every $8$ OFDM symbols within a single physical resource block (PRB).}
  \label{fig:PRS_Structure_Downsampled}
\end{figure}

During sensing snapshots, we set $\mathbf{X}=\mathbf{X}_p$, where $\mathbf{X}_p$ is a PRS-like pilot allocated on a resource-element (RE) set $\Omega$ (cf.~Fig.~\ref{fig:PRS_Structure_Downsampled}). Then, on PRS REs, the least-squares (LS) CFR estimate is
\begin{equation}
\tilde{\mathbf{H}}_p(n_{r_y},n_{r_z},n_{sc},l)=\frac{\mathbf{Y}(n_{r_y},n_{r_z},n_{sc},l)}{\mathbf{X}_p(n_{sc},l)},
\quad (n_{sc},l)\in\Omega.
\label{eq:ls_est}
\end{equation}

\subsection{DDAE Feature Tensor}
\label{sec:ddae_features}

% as in our prior

For each TX--RX link, we first obtain CFR snapshots from pilot-based LS estimation.
Instead of performing classical clutter suppression and CFAR/peak-picking as in 
model-based sensing pipeline, we treat the
Delay--Doppler--Azimuth--Elevation (DDAE) transform as a deterministic feature front-end.
The resulting 4-D tensor is converted to a log-power representation, quantized to
8-bit unsigned integers, and fed to a learned encoder. The encoder produces a compact
latent code that is transmitted to the Sensing Function (SF) for downstream AI/ML
inference.

Let $\tilde{H}(n_y,n_z,n_{sc},l)$ denote the complex CFR for an $N_{r_y}\times N_{r_z}$ RX UPA
over $N_{sc}$ subcarriers and $L$ OFDM symbols, with indices
$n_{sc}\in\{0,\dots,N_{sc}-1\}$ and $l\in\{0,\dots,L-1\}$.
Under the far-field plane-wave assumption, each propagation component is approximately
separable across frequency (delay), slow-time (Doppler), and space (AoA), so a
separable FFT/IFFT chain concentrates each path’s energy into a sparse neighborhood in a
4-D Fourier grid (delay, Doppler, azimuth, elevation).

\emph{Delay transform:} Although only $N_{\rm sc}=3276$ active subcarriers are populated, the
delay transform is evaluated on an $N_{\tau}$-point delay grid using
zero-padding. In the experiments, $N_{\tau}=4096$. Hence the delay-grid
spacing is
\[
\Delta_{\rm del}=\frac{1}{N_{\tau}\Delta f},
\]
whereas the nominal bandwidth-limited delay resolution is approximately
$1/(N_{\rm sc}\Delta f)$. For $\Delta f=120$ kHz and
$N_{\tau}=4096$, this gives $\Delta_{\rm del}=2.0345$ ns/bin.
$d_n\in\{0,\dots,N_{\tau}-1\}$:
\begin{align}
H_D(n_y,n_z,l,d_n) \;&=\; \frac{1}{N_{\tau}}\sum_{n_{sc}=0}^{N_{\tau}-1}\tilde{H}(n_y,n_z,n_{sc},l) \nonumber\\
& \quad \times\, e^{j2\pi n_{sc}d_n/N_{\tau}}.
\label{eq:delay_ifft}
\end{align}
Retain only the first $D_0$ bins corresponding to the maximum unambiguous range:
$H_{D,\mathrm{win}} = H_D(:,:,:,d_n<D_0)$.

\emph{Doppler transform:} apply a length-$L$ FFT across OFDM symbols to obtain Doppler bins
$f_n\in\{-L/2,\dots,L/2-1\}$:
\begin{align}
H_{DD}(n_y,n_z,f_n,d_n) \;&=\; \sum_{l=0}^{L-1} H_{D,\mathrm{win}}(n_y,n_z,l,d_n) \nonumber\\
& \quad \times\, e^{-j2\pi lf_n/L}.
\label{eq:doppler_fft}
\end{align}
Optionally retain a centered window of $F_0$ Doppler bins around zero Doppler.

\emph{Azimuth/elevation transforms:} apply spatial FFTs across the $y$- and $z$-axes of the RX UPA are \\$H_{DDA}(az_n,n_z,f_n,d_n)$ = 
\begin{equation}
 \sum_{n_y=0}^{N_{r_y}-1} H_{DD}(n_y,n_z,f_n,d_n)\,
e^{-j2\pi n_y az_n/N_{r_y}}, \label{eq:az_fft}\\
\end{equation} and 
$H_{DDAE}(az_n,el_n,f_n,d_n)=$ 
\begin{equation}
\sum_{n_z=0}^{N_{r_z}-1} H_{DDA}(az_n,n_z,f_n,d_n)\,
e^{-j2\pi n_z el_n/N_{r_z}}. \label{eq:el_fft}
\end{equation}
The resulting tensor $H_{DDAE}\in\mathbb{C}^{N_{r_y}\times N_{r_z}\times F_0 \times D_0}$
is the DDAE cube.
% \cite{shankar_fr3_ddae_baseline}
In contrast to the earlier pipeline where quasi-static clutter was suppressed by mean subtraction
across frequency and slow-time prior to the DDAE transform,
we do \emph{not} apply explicit clutter removal here. Our target environments include
complex, time-varying multipath and moving interferers, for which fixed heuristics can
remove useful structure and introduce artifacts. Instead, clutter/background suppression
is learned implicitly by the encoder operating on the raw DDAE magnitude features.

We convert the complex DDAE cube to a real-valued feature tensor by log-power in dB:
\begin{equation}
P(a,e,v,d) \;=\; 10\log_{10}\!\left(\left|H_{DDAE}(az_n,el_n,f_n,d_n)\right|^2 \right),
\label{eq:power_db}
\end{equation}.

To reduce model complexity and storage, we uniformly quantize
$P(\cdot)$ to an 8-bit unsigned tensor. Using a fixed dynamic range
$[P_{\min},P_{\max}]$ (in dB), we clip and map to $\{0,\dots,255\}$:
\begin{equation}
Q \;=\; \mathrm{uint8}\!\left(
\mathrm{round}\!\left(
255\cdot
\frac{\mathrm{clip}(P,P_{\min},P_{\max})-P_{\min}}{P_{\max}-P_{\min}}
\right)\right).
\label{eq:uint8_quant}
\end{equation}
The quantized tensor $Q$ is the input to the learned encoder. The encoder outputs a compact
latent representation that is transmitted to the SF, where downstream AI/ML models perform
detection/estimation.

% {architecture_tmlcn_single_target_modf.pdf}

\begin{figure*}[!t]
    \centering
    \includegraphics[
      width=1\textwidth,
      height=1\textheight,
      keepaspectratio,
      trim={0 0 0 0},
      clip]{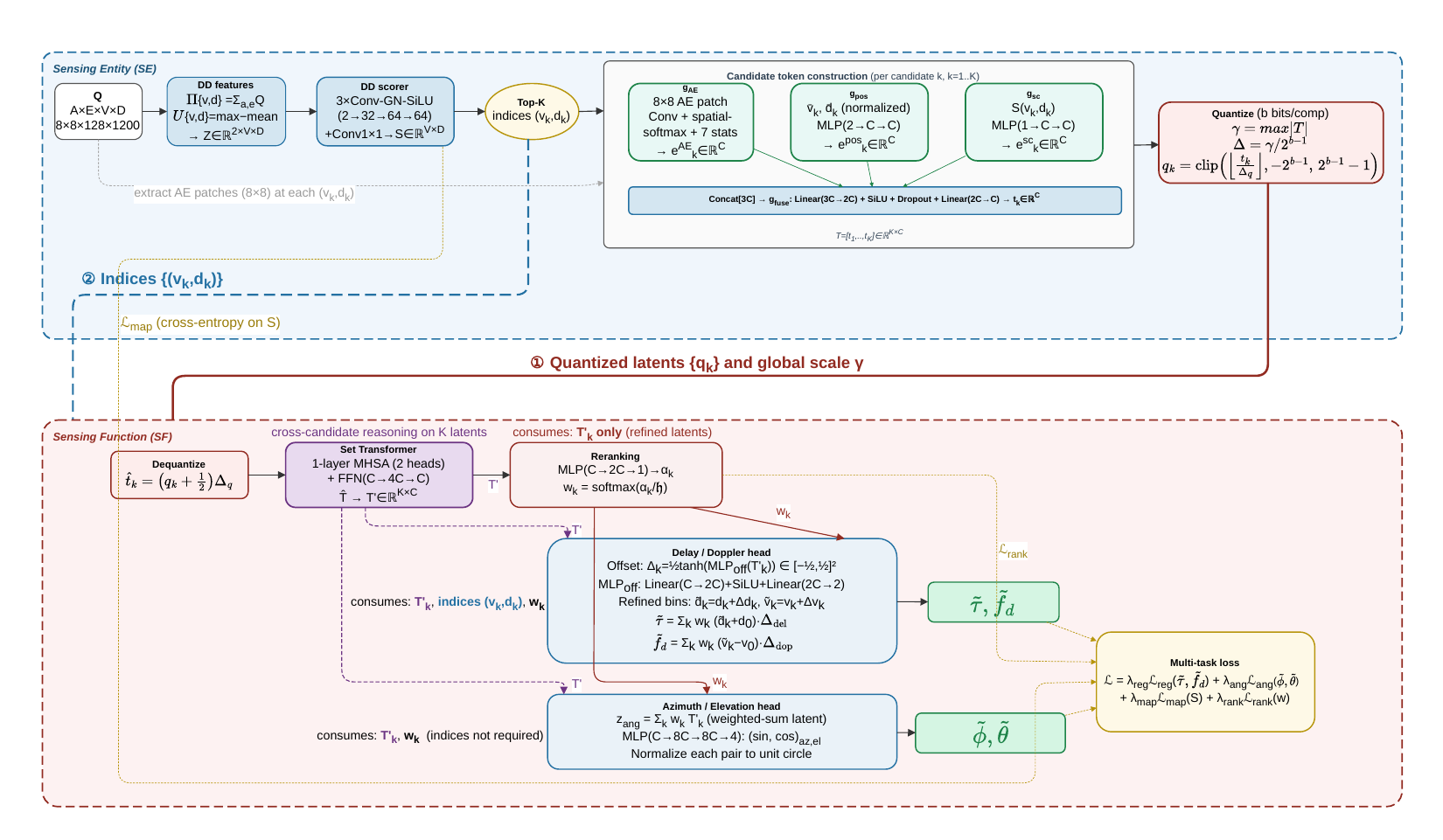}
    % \caption{Overview of the proposed coarse-to-fine SE--SF architecture. The SE constructs delay--Doppler features, scores candidates via a lightweight CNN, constructs candidate latents, and quantizes them for transmission. The SF receives the quantized latents, performs reranking via learned attention weights, and produces final delay, Doppler, azimuth, and elevation estimates.}
    \caption{End-to-end block diagram of the proposed coarse-to-fine ISAC sensing pipeline, showing SE-side candidate proposal and latent construction, the quantized feedback interface, and SF-side set refinement, reranking, and four-parameter estimation}
    \label{fig:architecture}
\end{figure*}

\section{Proposed Method}
\label{sec:method}

We take the quantized DDAE tensor $Q$ defined in
Sec.~\ref{sec:system_model} as input. As shown in Fig.~\ref{fig:architecture}, the
proposed pipeline comprises three stages that map cleanly onto the
SE--SF split: (i)~SE-side processing---candidate proposal and latent
construction (Sec.~\ref{subsec:se_side});
(ii)~a feedback interface---quantization, bit-budget accounting, and
dequantization (Sec.~\ref{subsec:feedback}); and (iii)~SF-side
processing---set-based latent refinement, reranking, and four-parameter
estimation from the received latents (Sec.~\ref{subsec:sf_side}).
The training objective is specified in Sec.~\ref{subsec:loss}.

% =============================================================================
\subsection{SE-Side Processing: Candidate Proposal and Latent Construction}
\label{subsec:se_side}
% =============================================================================

All operations in this subsection execute at the sensing entity. The SE
observes the DDAE tensor, generates a set of delay--Doppler candidate
hypotheses, and encodes each into a compact latent vector for
transmission.

\subsubsection{Delay--Doppler Feature Descriptor}
\label{subsubsec:dd_descriptor}

For each delay--Doppler bin $(v,d)$, we aggregate the angular response
map $Q_{:,:,v,d} \in \mathbb{R}^{A \times E}$ into two scalar
statistics: the total power
\begin{equation}
\Pi(v,d) \;=\; \sum_{a=1}^{A}\sum_{e=1}^{E} Q(a,e,v,d),
\label{eq:power}
\end{equation}
and the angular concentration (peakedness)
\begin{equation}
U(v,d) \;=\; \max_{a,e} Q(a,e,v,d)
             \;-\; \frac{1}{AE}\sum_{a=1}^{A}\sum_{e=1}^{E} Q(a,e,v,d).
\label{eq:peakedness}
\end{equation}
Stacking yields the two-channel descriptor
$\mathbf{Z} \in \mathbb{R}^{2 \times V \times D}$ with
$\mathbf{Z}_1 = \Pi$ and $\mathbf{Z}_2 = U$. The power channel highlights
bins with significant energy integrated over angle, while the
peakedness channel emphasizes bins whose energy is concentrated in a
small angular region---a signature more consistent with a coherent
target return than with diffuse clutter. Both are parameter-free
reductions and incur negligible overhead.

\subsubsection{Dense Delay--Doppler Scoring Network}
\label{subsubsec:scorer}

The descriptor $\mathbf{Z}$ is mapped to a per-bin confidence map by a
lightweight fully-convolutional network $\phi_{\theta}(\cdot)$. We
apply $L_s$ blocks of $3\!\times\!3$ convolution with group
normalization and SiLU activation, preserving spatial resolution,
followed by a $1\!\times\!1$ projection to a scalar confidence per
bin:
\begin{align}
\mathbf{H}^{(\ell)}
  &= \SiLU\!\left(\GN\!\left(
     \Conv_{3 \times 3}^{(\ell)}\!\left(
     \mathbf{H}^{(\ell-1)}\right)\right)\right),
     \;\; \ell = 1,\ldots,L_s,
     \label{eq:scorer_block}\\[2pt]
\mathbf{S} &= \Conv_{1 \times 1}\!\left(
              \mathbf{H}^{(L_s)}\right) \;\in\; \mathbb{R}^{V \times D},
     \label{eq:scorer_out}
\end{align}
with $\mathbf{H}^{(0)} = \mathbf{Z}$. The output $\mathbf{S}$ provides
unnormalized logits for all $V\!\times\!D$ delay--Doppler bins and is
supervised via a cross-entropy loss (Sec.~\ref{subsec:loss}) to
promote high recall of the ground-truth bin.

\subsubsection{Top-\texorpdfstring{$K$}{K} Candidate Selection}
\label{subsubsec:topk}

We vectorize
$\mathbf{s} = \operatorname{vec}(\mathbf{S}) \in \mathbb{R}^{VD}$ and
select the index set $\mathcal{I}_K = \TopK(\mathbf{s}, K)$.
Each flat index $i_k$ is mapped to two-dimensional coordinates via
% \begin{equation}
% v_k \;=\; \left\lfloor \tfrac{i_k - 1}{D} \right\rfloor + 1,
% \qquad
% d_k \;=\; i_k - v_k D,
% \label{eq:topk_unravel}
% \end{equation}
\begin{equation}
v_k=\left\lfloor\frac{i_k}{D}\right\rfloor,\qquad
d_k=i_k-v_kD .
\label{eq:topk_unravel}
\end{equation}
yielding the candidate set
$\mathcal{C} = \{(v_k, d_k, s_{i_k})\}_{k=0}^{K-1}$.
Because downstream modules at the SF cannot recover a hypothesis
absent from $\mathcal{C}$, the scorer is trained to prioritize recall
at~$K$. The value of $K$ is selected empirically in
Sec.~\ref{sec:results} to balance end-to-end detection rate
against the feedback payload.

\subsubsection{Candidate Token Construction}
\label{subsubsec:token}

For each candidate $(v_k, d_k) \in \mathcal{C}$, we construct a
$C$-dimensional latent that summarizes (i)~the angular structure at
the candidate bin, (ii)~the candidate position, and (iii)~the
candidate confidence. Specifically, we extract
\begin{itemize}
\item \emph{Angular patch:}
      $\mathrm{AE}_k = Q_{:,:,v_k,d_k} \in \mathbb{R}^{A \times E}$,
      the full azimuth--elevation response at the selected bin.
\item \emph{Normalized position:}
      $\bar{\mathbf{p}}_k = [\bar{v}_k, \bar{d}_k]^{\!\top}
      \in [-1,1]^2$, with $\bar{v}_k = 2v_k/(V-1) - 1$ and
      $\bar{d}_k = 2d_k/(D-1) - 1$.
\item \emph{Score:} $s_k \in \mathbb{R}$, the confidence logit from
      the scoring network.
\end{itemize}

Three lightweight encoders map these cues into a common
$C$-dimensional embedding space.

\paragraph*{Angular encoder $g_{\mathrm{AE}}(\cdot)$}
The $A \times E$ angular patch is first passed through a small
convolutional block (two $3\!\times\!3$ Conv--GN--SiLU layers at
hidden width~$h$) and projected into a scalar per-pixel score map via
a $1\!\times\!1$ convolution. A spatial softmax over the $A \times E$
grid converts this score map into a probability distribution
$p(a,e) \geq 0$, $\sum_{a,e} p(a,e) = 1$. From $p$ we compute the
expected angular coordinates and their spread on a fixed coordinate
grid $(y(a), z(e)) \in [-1,1]^2$ aligned with the UPA axes,
\begin{align}
\bar{y}
  &= \sum_{a,e} p(a,e)\, y(a),
\qquad
\bar{z}
   = \sum_{a,e} p(a,e)\, z(e),
\label{eq:sp_softmax_mean}\\[2pt]
\sigma_y^2
  &= \sum_{a,e} p(a,e)\,(y(a) - \bar{y})^2,
\nonumber\\
\sigma_z^2
  &= \sum_{a,e} p(a,e)\,(z(e) - \bar{z})^2.
\label{eq:sp_softmax_var}
\end{align}
Three scalar summaries of the raw patch---its peak value, mean, and
standard deviation---are appended, producing a seven-dimensional
feature vector
\begin{equation}
\mathbf{f}_k^{\mathrm{AE}}
= \big[\bar{y},\,\bar{z},\,\sigma_y^2,\,\sigma_z^2,\,
       \max \mathrm{AE}_k,\, \overline{\mathrm{AE}}_k,\,
       \sigma(\mathrm{AE}_k)\big]^{\!\top}
\in \mathbb{R}^{7},
\label{eq:ae_feat}
\end{equation}
which is mapped to $\mathbb{R}^C$ by a two-layer multi-layer perceptron (MLP) with SiLU
activation:
\begin{equation}
\mathbf{e}_k^{\mathrm{AE}}
\;=\; g_{\mathrm{AE}}(\mathrm{AE}_k)
\;=\; \mathbf{W}_2^{\mathrm{AE}}\,
      \SiLU\!\big(\mathbf{W}_1^{\mathrm{AE}}\,
      \mathbf{f}_k^{\mathrm{AE}}\big)
\;\in\; \mathbb{R}^C.
\label{eq:gae}
\end{equation}
The spatial-softmax readout yields a differentiable sub-pixel estimate
of the angular peak location, while the appended variance and
intensity statistics carry information about peak sharpness and
overall return strength. Together these cues are sufficient for the
SF-side angle head to recover azimuth and elevation from the
transmitted latent alone, without requiring the raw patch to cross
the feedback interface.

\paragraph*{Position and score encoders}
The position encoder $g_{\mathrm{pos}}(\cdot)$ is a two-layer MLP that
maps $\bar{\mathbf{p}}_k \in \mathbb{R}^2$ to $\mathbb{R}^C$,
\begin{equation}
\mathbf{e}_k^{\mathrm{pos}}
\;=\; \mathbf{W}_2^{\mathrm{pos}}\,
      \SiLU\!\big(\mathbf{W}_1^{\mathrm{pos}}\,
      \bar{\mathbf{p}}_k\big),
\label{eq:gpos}
\end{equation}
providing the fusion network with differentiable access to the
candidate's normalized delay--Doppler coordinates. The score encoder
$g_{\mathrm{sc}}(\cdot)$ is a two-layer MLP that maps
$s_k \in \mathbb{R}$ to $\mathbb{R}^C$,
\begin{equation}
\mathbf{e}_k^{\mathrm{sc}}
\;=\; \mathbf{W}_2^{\mathrm{sc}}\,
      \SiLU\!\big(\mathbf{W}_1^{\mathrm{sc}}\,s_k\big),
\label{eq:gsc}
\end{equation}
exposing the proposal-stage confidence as an additional discriminative
feature for candidate fusion.

\paragraph*{Fusion into a candidate token}
The three embeddings are concatenated into a $3C$-dimensional vector
and mapped to a $C$-dimensional candidate token by a two-layer MLP
with dropout,
\begin{equation}
\mathbf{t}_k
= \mathbf{W}_2^{\mathrm{fuse}}\,
  \mathrm{Dropout}\!\Big(
  \SiLU\!\big(
  \mathbf{W}_1^{\mathrm{fuse}}
  [\mathbf{e}_k^{\mathrm{AE}};\,
   \mathbf{e}_k^{\mathrm{pos}};\,
   \mathbf{e}_k^{\mathrm{sc}}]\big)\Big),
\label{eq:fuse}
\end{equation}
with $\mathbf{W}_1^{\mathrm{fuse}} \in \mathbb{R}^{2C \times 3C}$ and
$\mathbf{W}_2^{\mathrm{fuse}} \in \mathbb{R}^{C \times 2C}$. Stacking
$\{\mathbf{t}_k\}_{k=1}^{K}$ yields the candidate latent matrix
$\mathbf{T} = [\mathbf{t}_1, \ldots, \mathbf{t}_K]^{\!\top}
\in \mathbb{R}^{K \times C}$, which is the SE's output.

\medskip
\noindent\textbf{Remark (SE--SF boundary).}
At this point, the SE has completed all local computation. The
candidate latent matrix $\mathbf{T}$, the candidate indices
$\{(v_k, d_k)\}_{k=1}^{K}$, and a single global scale parameter
$\gamma_q$ together constitute the information transmitted to the SF.
No raw angular patch, raw DDAE slice, or channel-estimate tensor is
forwarded; the SF must perform all downstream inference from the
transmitted latents and indices alone. The next subsection formalizes
this feedback interface and its bit-budget accounting.

% =============================================================================
\subsection{SE--SF Feedback Interface: Quantization and Bit Budget}
\label{subsec:feedback}
% =============================================================================

\subsubsection{Bit Budget}
\label{subsubsec:budget}

The feedback message comprises three components transmitted from the
SE to the SF: (i)~the quantized candidate latents
$\{\mathbf{q}_k\}_{k=1}^{K}$; (ii)~the corresponding delay--Doppler
candidate indices $\{(v_k, d_k)\}_{k=1}^{K}$; and (iii)~quantizer side
information consisting of a global per-sample scale $\gamma_q$. With
$b$ bits per latent component, $K$ candidates of dimension $C$, and a
delay--Doppler grid of size $VD$, the total feedback budget is
\begin{equation}
B_{\mathrm{fb}}
\;=\;
\underbrace{KCb}_{\text{quantized latents}}
\;+\;
\underbrace{K \lceil \log_2(VD) \rceil}_{\text{candidate indices}}
\;+\;
\underbrace{B_{\mathrm{side}}}_{\text{global scale}},
\label{eq:budget_general}
\end{equation}
where a single global scale parameter
($B_{\mathrm{side}} = 16$~bits in our implementation) is broadcast
once per sample. Per-candidate scales would replace $B_{\mathrm{side}}$
by $K B_{\mathrm{side}}$ but were not found beneficial in our
experiments. Equation~\eqref{eq:budget_general} enables direct
payload--accuracy evaluation by sweeping $b$ at fixed $(K,C)$.

For a delay--Doppler grid of size $V \times D = 128 \times 1200$, we
have $\lceil \log_2(153\,600) \rceil = 18$ bits per candidate index,
so the feedback budget reduces to
\begin{equation}
B_{\mathrm{fb}} \;=\; bKC \;+\; 18K \;+\; 16 ~\text{bits}.
\label{eq:budget_selected}
\end{equation}
For the standard (STD) anchor $(K, C, b) = (8, 64, 6)$ used as a
running example in this paper, the latent payload is $3{,}072$~bits,
the candidate-index overhead is $144$~bits, and the global scale
contributes $16$~bits, totaling $3{,}232$~bits or $404$~bytes per
sample. The same expression evaluated at the ultra-compact (UC)
anchor $(4, 32, 6)$ and the high-fidelity (HF) anchor $(16, 64, 6)$
gives $107$~B and $806$~B per sample, respectively. This accounting
is bit-exact and underlies all payload numbers reported in
Sec.~\ref{sec:results}. Halving the candidate count~$K$ halves both
the latent and the index contributions simultaneously, enabling
aggressive low-bandwidth operating points, which we study in
Sec.~\ref{sec:quant_sensitivity}.

\subsubsection{Uniform Mid-Rise Quantizer and Dequantizer}
\label{subsubsec:midrise}

We employ a symmetric mid-rise uniform quantizer with $2^b$
reconstruction levels and no level placed at zero. Given a clipping
threshold $\gamma_q > 0$ and a bit budget $b \geq 1$ per latent
component, the step size is
\begin{equation}
\Delta_q \;=\; \frac{\gamma_q}{2^{\,b-1}},
\label{eq:qstep}
\end{equation}
and the $2^b$ levels are placed symmetrically about the origin at
\begin{equation}
r_q \;=\; \left(q + \tfrac{1}{2}\right)\Delta_q,
\quad
q \in \mathcal{Q}_b \triangleq \{-2^{b-1},\ldots,2^{b-1} - 1\}.
\label{eq:qlevels}
\end{equation}
This assigns $2^{b-1}$ levels per sign and a uniform spacing
$\Delta_q$ between consecutive reconstruction values at every bit
budget. We use $\Delta_q$ here to distinguish the quantizer step from
the Doppler-bin and delay-bin resolutions $\Delta_{\mathrm{dop}}$ and
$\Delta_{\mathrm{del}}$ used in Sec.~\ref{subsec:sf_side}. For the
lowest bit budgets, \eqref{eq:qlevels} reduces to
\begin{align*}
b=1: &\;\; r_q \in \{-\tfrac{1}{2},+\tfrac{1}{2}\}\gamma_q,
      \;\;\; \Delta_q = \gamma_q,\\
b=2: &\;\; r_q \in \{-\tfrac{3}{4},-\tfrac{1}{4},
           +\tfrac{1}{4},+\tfrac{3}{4}\}\gamma_q,
      \;\;\; \Delta_q = \gamma_q/2,\\
b=3: &\;\; r_q \in \{\pm\tfrac{1}{8},\pm\tfrac{3}{8},
           \pm\tfrac{5}{8},\pm\tfrac{7}{8}\}\gamma_q,
      \;\;\; \Delta_q = \gamma_q/4,
\end{align*}
and analogously for higher $b$. The 1-bit case coincides with scaled
sign quantization,
$\hat{t} = \tfrac{\gamma_q}{2}\,\sgn(t)$.

\paragraph*{Encoding (SE side)}
Each latent component $t \in \mathbb{R}$ of $\mathbf{t}_k$ is clipped
to $[-\gamma_q, \gamma_q]$ and mapped to its quantizer index by
\begin{equation}
q(t) \;=\; \operatorname{clip}\!\left(
          \left\lfloor \tfrac{t}{\Delta_q} \right\rfloor,\,
          -2^{b-1},\, 2^{b-1} - 1\right),
\label{eq:qencode}
\end{equation}
applied component-wise to obtain $\mathbf{q}_k = q(\mathbf{t}_k) \in
\mathcal{Q}_b^{C}$. Under the bijection
$q \mapsto q + 2^{b-1} \in \{0, \ldots, 2^b - 1\}$ used for
transmission, each candidate latent occupies exactly $Cb$~bits. The
clipping threshold is set per sample from the empirical dynamic range
of the latent matrix,
\begin{equation}
\gamma_q \;=\; \max_{1 \leq k \leq K,\; 1 \leq c \leq C}
           \big|\mathbf{T}_{k,c}\big|,
\label{eq:gamma_def}
\end{equation}
and transmitted as a 16-bit floating-point side-information parameter.

\paragraph*{Decoding (SF side)}
The SF reconstructs each component via \eqref{eq:qlevels} as
\begin{equation}
\hat{t} \;=\; \left(q + \tfrac{1}{2}\right)\Delta_q,
\label{eq:qdecode}
\end{equation}
yielding $\widehat{\mathbf{T}} = [\hat{\mathbf{t}}_1, \ldots,
\hat{\mathbf{t}}_K]^{\!\top} \in \mathbb{R}^{K \times C}$. For
unsaturated inputs, the worst-case reconstruction error is bounded by
$\Delta_q / 2 = \gamma_q / 2^{\,b}$, which halves with each additional
bit; saturation error is controlled by $\gamma_q$.

\paragraph*{Implementation and baselines}
We also report two floating-point baselines in which each latent
component is transmitted in FP32 (32~bits) or FP16 (16~bits). In our
implementation, the quantizer--dequantizer pair
\eqref{eq:qencode}--\eqref{eq:qdecode} is inserted \emph{post-training}
during inference, so that the SF-side estimator operates on
$\widehat{\mathbf{T}}$ with no change to the trained weights. This
isolates the effect of feedback compression from training dynamics.
The same interface supports quantization-aware training by treating
\eqref{eq:qencode} as the identity in the backward pass
(straight-through estimator) when further robustness to very low bit
budgets is required; exploring such quantization-aware variants is
left to future work.

% =============================================================================
\subsection{SF-Side Processing: Set Refinement, Reranking, and Estimation}
\label{subsec:sf_side}
% =============================================================================

All operations in this subsection execute at the sensing function
using only the received quantities: the reconstructed latent matrix
$\widehat{\mathbf{T}} = [\hat{\mathbf{t}}_1, \ldots,
\hat{\mathbf{t}}_K]^{\!\top} \in \mathbb{R}^{K \times C}$ and the
candidate indices $\{(v_k, d_k)\}_{k=1}^{K}$. The SF first refines
the candidate latents through a set-based attention block that allows
cross-candidate reasoning, then computes a soft weighting over the
refined hypotheses, and finally uses these weights for delay, Doppler,
azimuth, and elevation estimation.

\subsubsection{Set-Based Latent Refinement}
\label{subsubsec:settf}

The received candidates form an unordered set whose elements are
coupled through the physical scene: multiple candidates may belong to
the same target neighborhood, clutter responses may share
characteristic angular signatures, and the relative confidences of
nearby candidates carry joint information that a candidate-independent
head cannot exploit. To capture these interactions, we pass
$\widehat{\mathbf{T}}$ through a permutation-equivariant self-attention
block $f_{\eta}(\cdot)$,
\begin{equation}
\mathbf{T}' \;=\; f_{\eta}\!\left(\widehat{\mathbf{T}}\right)
\;\in\; \mathbb{R}^{K \times C},
\label{eq:settf_out}
\end{equation}
% implemented as a single Transformer layer with multi-head
% self-attention (MHSA), layer normalization (LN), and a position-wise
% feed-forward network (FFN) of expansion ratio four. Concretely,
% \begin{align}
% \mathbf{U} &= \LN\!\left(
%               \widehat{\mathbf{T}}
%               \;+\; \MHSA\!\left(\widehat{\mathbf{T}}\right)\right),
%    \label{eq:settf_mhsa}\\[2pt]
% \mathbf{T}' &= \LN\!\left(
%                \mathbf{U} \;+\; \FFN(\mathbf{U})\right),
%    \label{eq:settf_ffn}
% \end{align}
% with
% \begin{equation}
% \FFN(\mathbf{u})
% \;=\; \mathbf{W}_4^{\mathrm{ff}}\,
%       \SiLU\!\left(
%       \mathbf{W}_3^{\mathrm{ff}}\,\mathbf{u}\right),
% \label{eq:settf_ffn_def}
% \end{equation}
% $\mathbf{W}_3^{\mathrm{ff}} \in \mathbb{R}^{4C \times C}$ and
% $\mathbf{W}_4^{\mathrm{ff}} \in \mathbb{R}^{C \times 4C}$. The block
% is permutation-equivariant: reordering the input candidates produces
% a correspondingly reordered output, so no positional encoding is
% applied and no ordering is imposed on the candidate set. Each output
% vector $\mathbf{t}'_k$ is a context-aware refinement of
% $\hat{\mathbf{t}}_k$ that reflects the local structure of the full
% set, enabling the reranker below to suppress candidates whose latents
% closely resemble nearby ones of higher confidence and to emphasize
% candidates whose latents are distinctive.
implemented as a single Transformer layer with multi-head
self-attention (MHSA), pre-norm layer normalization (LN), and a
position-wise feed-forward network (FFN) of expansion ratio four.
We adopt the pre-norm convention, applying LN to each sublayer
input with the residual added afterwards, which improves training
stability without learning-rate warmup~\cite{xiong2020layernorm}.
The block uses 2 attention heads (head dimension $d_h = C/2$)
and dropout $p=0.1$ applied to the attention weights and the FFN
output. Concretely,
\begin{align}
\mathbf{U} &= \widehat{\mathbf{T}} + \mathrm{MHSA}\!\left(\mathrm{LN}(\widehat{\mathbf{T}})\right), \label{eq:settf_attn} \\
\mathbf{T}' &= \mathbf{U} + \mathrm{FFN}\!\left(\mathrm{LN}(\mathbf{U})\right), \label{eq:settf_ffn}
\end{align}
with
\begin{equation}
\mathrm{FFN}(\mathbf{u}) \;=\; \mathbf{W}^{\mathrm{ff}}_{4}\,\mathrm{SiLU}\!\left(\mathbf{W}^{\mathrm{ff}}_{3}\,\mathbf{u}\right),
\label{eq:settf_ffndef}
\end{equation}
$\mathbf{W}^{\mathrm{ff}}_{3} \in \mathbb{R}^{4C \times C}$ and
$\mathbf{W}^{\mathrm{ff}}_{4} \in \mathbb{R}^{C \times 4C}$. No
additional normalization is applied after the block; the output
$\mathbf{T}'$ is passed directly to the reranker and estimation
heads.

\subsubsection{Candidate Reranking}
\label{subsubsec:rerank}

The SF computes a soft weighting over the $K$ refined candidates by
mapping each to a scalar logit through a shared two-layer MLP
$r_{\psi}(\cdot)$,
\begin{equation}
\alpha_k \;=\; \mathbf{W}_6^{\mathrm{rr}}\,
               \SiLU\!\left(
               \mathbf{W}_5^{\mathrm{rr}}\,\mathbf{t}'_k\right),
\qquad k = 1,\ldots,K,
\label{eq:rerank_logit}
\end{equation}
with $\mathbf{W}_5^{\mathrm{rr}} \in \mathbb{R}^{2C \times C}$ and
$\mathbf{W}_6^{\mathrm{rr}} \in \mathbb{R}^{1 \times 2C}$, and
normalizing with a temperature-scaled softmax,
\begin{equation}
w_k \;=\; \frac{\exp(\alpha_k / \beta)}
               {\sum_{j=1}^{K} \exp(\alpha_j / \beta)},
\quad
\sum_{k=1}^{K} w_k = 1,\; w_k \geq 0,
\label{eq:rerank_softmax}
\end{equation}
where $\beta > 0$ is the temperature. The weights
$\mathbf{w} = [w_1, \ldots, w_K]^{\!\top}$ quantify the relative
plausibility of each candidate conditioned on the refined latents and
are shared across all downstream estimation heads. The ablation study
in Sec.\!\!~\ref{sec:results}-\!\!~\ref{sec:ablation_studies} shows that this reranking step is the
single most important component of the SF: removing it causes the
detection rate to drop by more than twenty percentage points, even
though the ground-truth candidate remains in the Top-$K$ set in over
$99\%$ of cases.

\subsubsection{Delay and Doppler Estimation}
\label{subsubsec:delay_doppler}

Each candidate coordinate $(v_k, d_k)$ is augmented with a learned
sub-bin offset predicted from $\mathbf{t}'_k$. An offset head
$o_{\eta}(\cdot)$ produces a bounded refinement,
\begin{equation}
\boldsymbol{\Delta}_k
\;=\;
\begin{bmatrix}\Delta v_k\\ \Delta d_k\end{bmatrix}
\;=\;
\frac{1}{2}\tanh\!\left(
\mathbf{W}_8^{\mathrm{off}}\,\SiLU\!\left(
\mathbf{W}_7^{\mathrm{off}}\,\mathbf{t}'_k\right)\right),
\label{eq:offset}
\end{equation}
with $\mathbf{W}_7^{\mathrm{off}} \in \mathbb{R}^{2C \times C}$,
$\mathbf{W}_8^{\mathrm{off}} \in \mathbb{R}^{2 \times 2C}$, and
$\Delta v_k, \Delta d_k \in [-\tfrac{1}{2},\tfrac{1}{2}]$. The bounded
refinement guarantees that each refined coordinate remains within its
own bin cell, preventing candidates from crossing bin boundaries and
avoiding multimodal ambiguity in the subsequent weighted sum. The
refined coordinates are
\begin{equation}
\tilde{v}_k = v_k + \Delta v_k,\qquad
\tilde{d}_k = d_k + \Delta d_k.
\label{eq:refined_coords}
\end{equation}
Using the Doppler-bin resolution $\Delta_{\mathrm{dop}}$ (Hz/bin) and
delay-bin resolution $\Delta_{\mathrm{del}}$ (s/bin), the
corresponding physical values are
\begin{equation}
\tilde{f}_{d,k}
= (\tilde{v}_k - v_0)\,\Delta_{\mathrm{dop}},
\quad
\tilde{\tau}_k
= (\tilde{d}_k + d_0)\,\Delta_{\mathrm{del}},
\label{eq:phys_values}
\end{equation}
where $v_0$ is the zero-Doppler bin index and $d_0$ is the starting delay-bin offset of the retained window. The final delay and
Doppler estimates are weighted sums over candidates,
\begin{equation}
\tilde{f}_d
= \sum_{k=1}^{K} w_k\,\tilde{f}_{d,k},
\qquad
\tilde{\tau}
= \sum_{k=1}^{K} w_k\,\tilde{\tau}_k.
\label{eq:final_dd}
\end{equation}
When the weight distribution is sharp---as occurs at convergence,
where the reranker typically places most of its mass on a single
candidate---this behaves as soft selection of the most plausible
hypothesis. For ambiguous scenes, it interpolates among nearby
candidates in a differentiable manner.

\subsubsection{Azimuth and Elevation Estimation}
\label{subsubsec:ang_estimation}

Rather than reconstructing a focused angular map at the SF, we form a
latent-level angular summary directly from the refined candidate
representations,
\begin{equation}
\mathbf{z}_{\mathrm{ang}}
\;=\;
\sum_{k=1}^{K} w_k\,\mathbf{t}'_k
\;\in\;\mathbb{R}^{C}.
\label{eq:zang}
\end{equation}
The vector $\mathbf{z}_{\mathrm{ang}}$ is a weighted fusion of the
latents that survive reranking; because $g_{\mathrm{AE}}$ was trained
to preserve the spatial-softmax peak location and spread in the
candidate latent, $\mathbf{z}_{\mathrm{ang}}$ concentrates the
target-relevant angular information available in the selected
candidates into a single $C$-dimensional representation. An angle
head $h_{\omega}(\cdot)$, implemented as a three-layer MLP with
hidden width $8C$,
\begin{equation}
h_{\omega}(\mathbf{z})
\;=\;
\mathbf{W}_{11}^{\mathrm{ang}}\,
\SiLU\!\left(
\mathbf{W}_{10}^{\mathrm{ang}}\,
\SiLU\!\left(
\mathbf{W}_{9}^{\mathrm{ang}}\,\mathbf{z}\right)\right),
\label{eq:anghead}
\end{equation}
with
$\mathbf{W}_{9}^{\mathrm{ang}} \in \mathbb{R}^{8C \times C}$,
$\mathbf{W}_{10}^{\mathrm{ang}} \in \mathbb{R}^{8C \times 8C}$, and
$\mathbf{W}_{11}^{\mathrm{ang}} \in \mathbb{R}^{4 \times 8C}$, then
predicts the sine and cosine components of both angles,
\begin{equation}
[\tilde{s}_{\mathrm{az}},\tilde{c}_{\mathrm{az}},
 \tilde{s}_{\mathrm{el}},\tilde{c}_{\mathrm{el}}]
= h_{\omega}(\mathbf{z}_{\mathrm{ang}}).
\label{eq:anghead_out}
\end{equation}
The output pairs are individually normalized to the unit circle,
\begin{equation}
\begin{bmatrix}\tilde{s}_{\mathrm{az}}\\ \tilde{c}_{\mathrm{az}}\end{bmatrix}
\leftarrow
\frac{[\tilde{s}_{\mathrm{az}},\tilde{c}_{\mathrm{az}}]^{\!\top}}
     {\sqrt{\tilde{s}_{\mathrm{az}}^{2} + \tilde{c}_{\mathrm{az}}^{2}
            + \epsilon}},
\label{eq:ang_unitnorm}
\end{equation}
and analogously for elevation, ensuring that the predictions lie on a
valid angle manifold. The final angle estimates are recovered by
\begin{equation}
\tilde{\phi}
= \mathrm{atan2}(\tilde{s}_{\mathrm{az}},\tilde{c}_{\mathrm{az}}),
\quad
\tilde{\theta}
= \mathrm{atan2}(\tilde{s}_{\mathrm{el}},\tilde{c}_{\mathrm{el}}).
\label{eq:final_ang}
\end{equation}
In this way, the SF performs angular estimation using only the
refined candidate latents and their reranker weights, without
requiring transmission of raw angular patches across the SE--SF
interface.

% =============================================================================
\subsection{Training Objective}
\label{subsec:loss}
% =============================================================================

The proposed network is trained end-to-end with a multi-task
objective that combines four components: (i)~continuous regression
for delay and Doppler; (ii)~angular supervision for azimuth and
elevation; (iii)~dense classification over the delay--Doppler grid to
promote high-recall candidate proposal; and (iv)~candidate-level
reranking supervision when a near-ground-truth hypothesis is present
in the Top-$K$ set.

Let the ground-truth target parameters be
$(\tau^{\star}, f_d^{\star}, \phi^{\star}, \theta^{\star})$, and let
the final estimates produced by the SF-side pipeline be
$(\tilde{\tau}, \tilde{f}_d, \tilde{\phi}, \tilde{\theta})$. Let
$\mathbf{S} \in \mathbb{R}^{V \times D}$ denote the dense
delay--Doppler logit map from the proposal network, and let
$\boldsymbol{\alpha} = [\alpha_1, \ldots, \alpha_K]^{\!\top}$ and
$\mathbf{w} = [w_1, \ldots, w_K]^{\!\top}$ denote the reranking logits
and soft weights defined in the candidate-reranking step of Sec.~\ref{subsec:sf_side}.

To supervise the proposal stage, the continuous ground-truth pair
$(\tau^{\star}, f_d^{\star})$ is mapped to its discretized
delay--Doppler bin $(d^{\star}, v^{\star})$, and the corresponding
flattened class index is
\begin{equation}
i^{\star} \;=\; v^{\star}\, D + d^{\star}.
\label{eq:gt_flat}
\end{equation}
The delay--Doppler regression loss is defined using a Smooth-$\ell_1$
penalty on normalized delay and Doppler errors,
\begin{equation}
\mathcal{L}_{\mathrm{reg}}
= \mathrm{Smooth\text{-}}\ell_1\!\left(
  \frac{\tilde{\tau} - \tau^{\star}}{s_{\tau}}\right)
+ \mathrm{Smooth\text{-}}\ell_1\!\left(
  \frac{\tilde{f}_d - f_d^{\star}}{s_f}\right),
\label{eq:L_reg}
\end{equation}
where $s_{\tau}$ and $s_f$ are fixed scale factors used to balance the
relative magnitudes of delay and Doppler errors in the combined loss.

For azimuth and elevation, we avoid wrap-around discontinuities by
supervising the sine/cosine outputs of the angle head rather than the
angles directly. Let
\begin{equation}
\mathbf{u}(\psi) = [\sin\psi,\;\cos\psi]^{\!\top}.
\label{eq:unitvec}
\end{equation}
Using the predicted sine/cosine pairs from~\eqref{eq:anghead_out},
the angular loss is
\begin{equation}
\mathcal{L}_{\mathrm{ang}}
=
\left\|
\begin{bmatrix}\tilde{s}_{\mathrm{az}}\\
               \tilde{c}_{\mathrm{az}}\end{bmatrix}
- \mathbf{u}(\phi^{\star})
\right\|_2^2
+
\left\|
\begin{bmatrix}\tilde{s}_{\mathrm{el}}\\
               \tilde{c}_{\mathrm{el}}\end{bmatrix}
- \mathbf{u}(\theta^{\star})
\right\|_2^2.
\label{eq:L_ang}
\end{equation}
This formulation is smooth and naturally respects the periodicity of
the angular variables.

The dense proposal map is supervised with a cross-entropy loss over
the vectorized delay--Doppler grid,
\begin{equation}
\mathcal{L}_{\mathrm{map}}
\;=\;
\CE\!\left(\operatorname{vec}(\mathbf{S}),\, i^{\star}\right),
\label{eq:L_map}
\end{equation}
which directly encourages the ground-truth delay--Doppler hypothesis
to receive a high score and thereby improves the recall of the Top-$K$
candidate set.

To supervise the reranking module, we first associate the ground
truth with the nearest retained candidate whenever the proposal stage
places a sufficiently close hypothesis in the Top-$K$ set. Let
\begin{equation}
\delta_k^{2}
= (v_k - v^{\star})^{2} + (d_k - d^{\star})^{2},
\quad
k^{\star} = \arg\min_k\, \delta_k^{2}.
\label{eq:nearest_cand}
\end{equation}
If $\delta_{k^{\star}}^{2} \leq \tau_{\mathrm{near}}^{2}$, where
$\tau_{\mathrm{near}}$ is a fixed neighborhood threshold, the
reranking loss is defined as
\begin{equation}
\mathcal{L}_{\mathrm{rank}}
\;=\;
\CE(\boldsymbol{\alpha},\, k^{\star}).
\label{eq:L_rank}
\end{equation}
Otherwise, the sample is excluded from reranking supervision, since
none of the retained candidates is sufficiently close to the
ground-truth bin.

The total training objective is
\begin{equation}
\mathcal{L}
\;=\;
\lambda_{\mathrm{reg}}\,\mathcal{L}_{\mathrm{reg}}
+ \lambda_{\mathrm{ang}}\,\mathcal{L}_{\mathrm{ang}}
+ \lambda_{\mathrm{map}}\,\mathcal{L}_{\mathrm{map}}
+ \lambda_{\mathrm{rank}}\,\mathcal{L}_{\mathrm{rank}},
\label{eq:total_loss}
\end{equation}
where $\lambda_{\mathrm{reg}}$, $\lambda_{\mathrm{ang}}$,
$\lambda_{\mathrm{map}}$, and $\lambda_{\mathrm{rank}}$ control the
relative strengths of the regression, angular, proposal, and
reranking terms. During training, the reranking weight
$\lambda_{\mathrm{rank}}$ is linearly warmed up from~$0$ to its
target value $\lambda_{\mathrm{rank}}^{\max}$ over the initial epochs
to stabilize optimization before the proposal stage becomes
sufficiently reliable. Soft teacher forcing is used during early
training by mixing the learned candidate weights with a
near-ground-truth one-hot supervision when such a candidate exists;
the mixing coefficient is annealed linearly from $1.0$ at the start
of training to~$0.0$ at the final epoch. All results reported in this
paper are obtained using the fully learned inference pipeline at test
time, without teacher forcing.

\section{Simulation Setup}
\label{sec:setup}

\begin{figure*}[!t]
    \centering
    \subfloat[\label{fig:sc_1}]{%
        \includegraphics[width=0.24\textwidth,trim={0 25 0 0},clip]{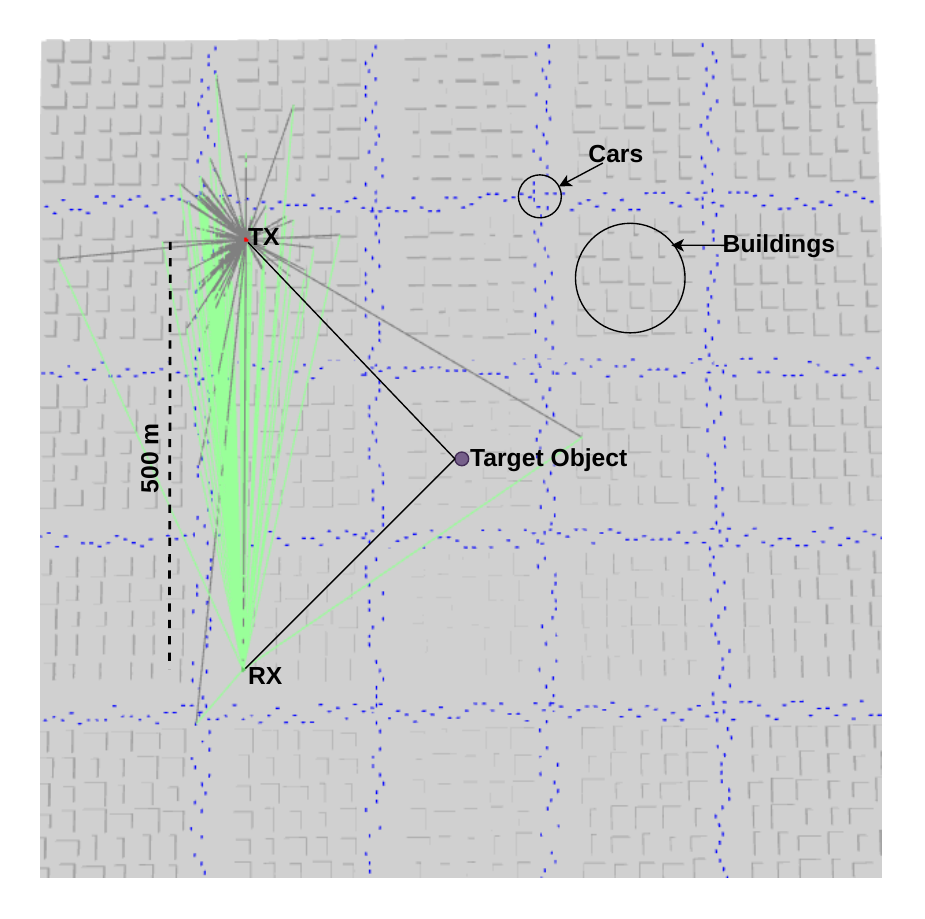}}
    \hfill
    \subfloat[\label{fig:sc_2}]{%
        \includegraphics[width=0.24\textwidth,trim={0 25 0 0},clip]{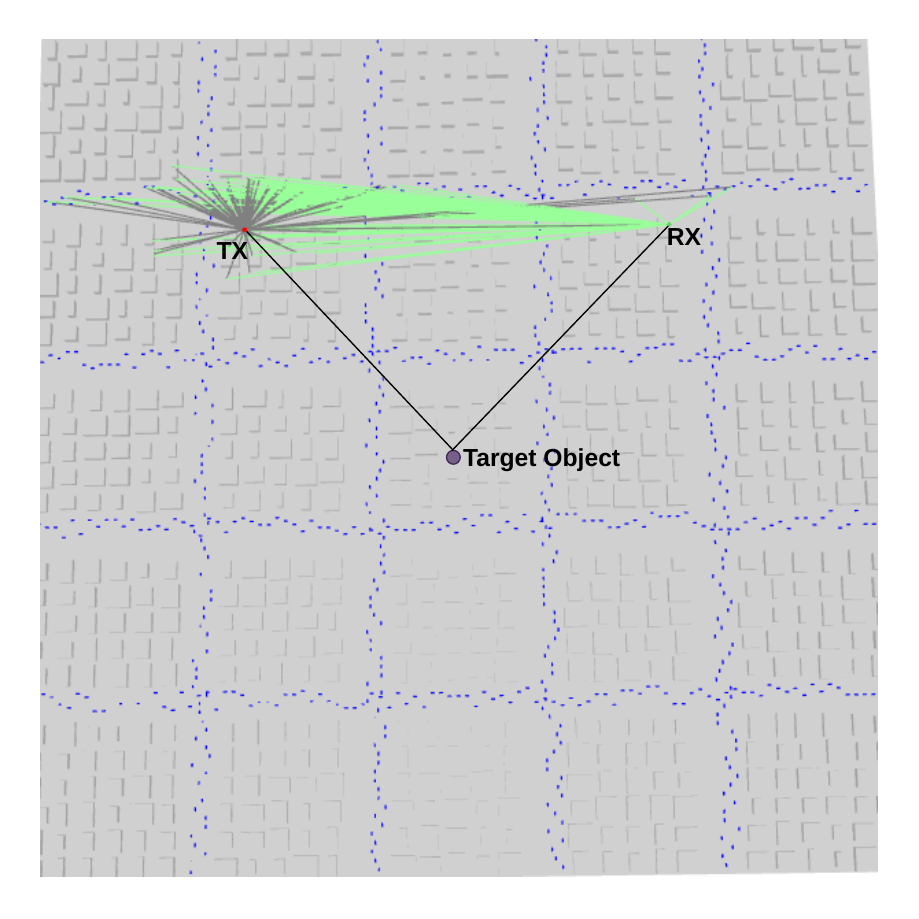}}
    \hfill
    \subfloat[\label{fig:sc_3}]{%
        \includegraphics[width=0.24\textwidth,trim={0 25 0 0},clip]{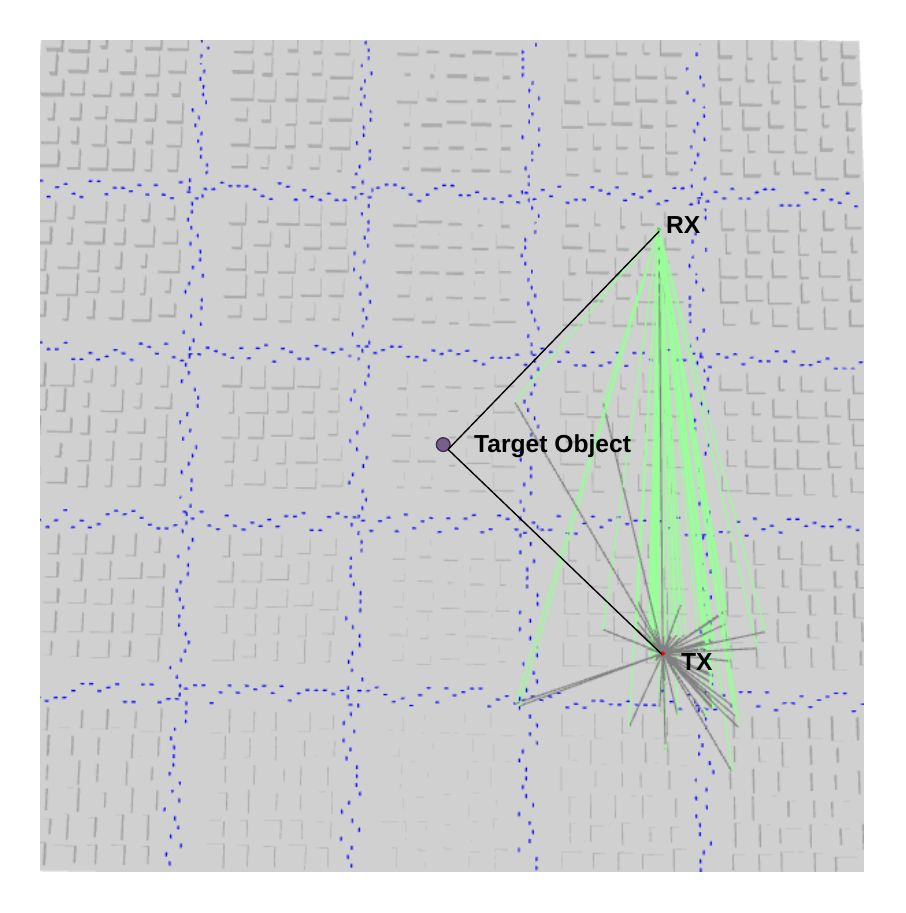}}
    \hfill
    \subfloat[\label{fig:sc_4}]{%
        \includegraphics[width=0.24\textwidth,trim={0 25 0 0},clip]{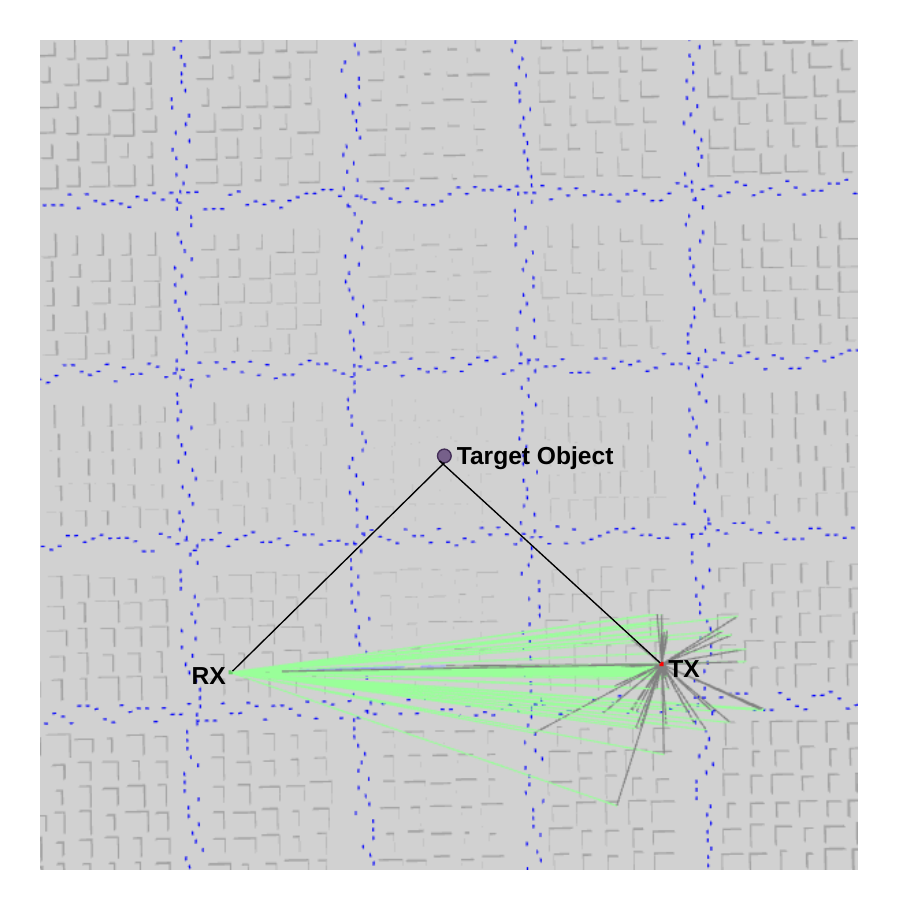}}
    \caption{Synthetic urban scene (Scene-Synth) with four bistatic TX--RX configurations. Buildings, roads, and vehicles are generated in Blender; ray tracing is performed with NVIDIA Sionna.}
    \label{fig:sc_all}
\end{figure*}

\begin{figure}[!t]
    \centering
    \subfloat[\label{fig:IITM_1}]{%
        \includegraphics[width=0.24\textwidth,trim={0 25 0 0},clip]{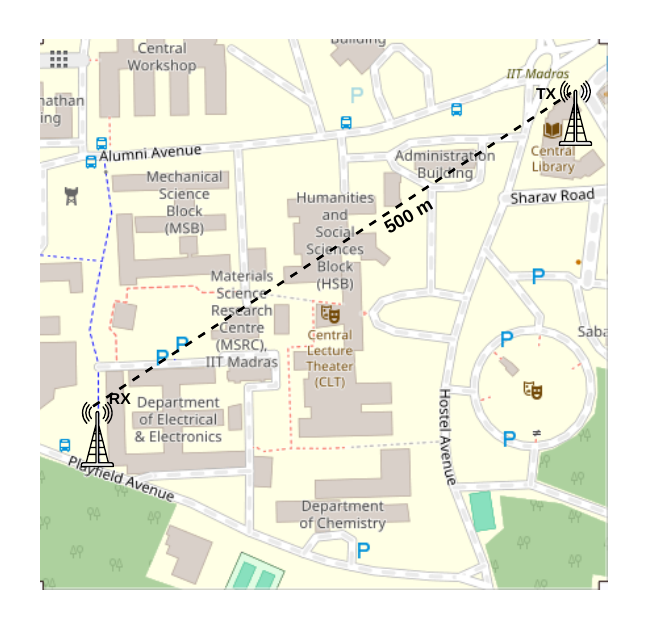}}
    \hfill
    \subfloat[\label{fig:IITM_2}]{%
        \includegraphics[width=0.24\textwidth,height=0.155\textheight, trim={0 30 0 0},clip]{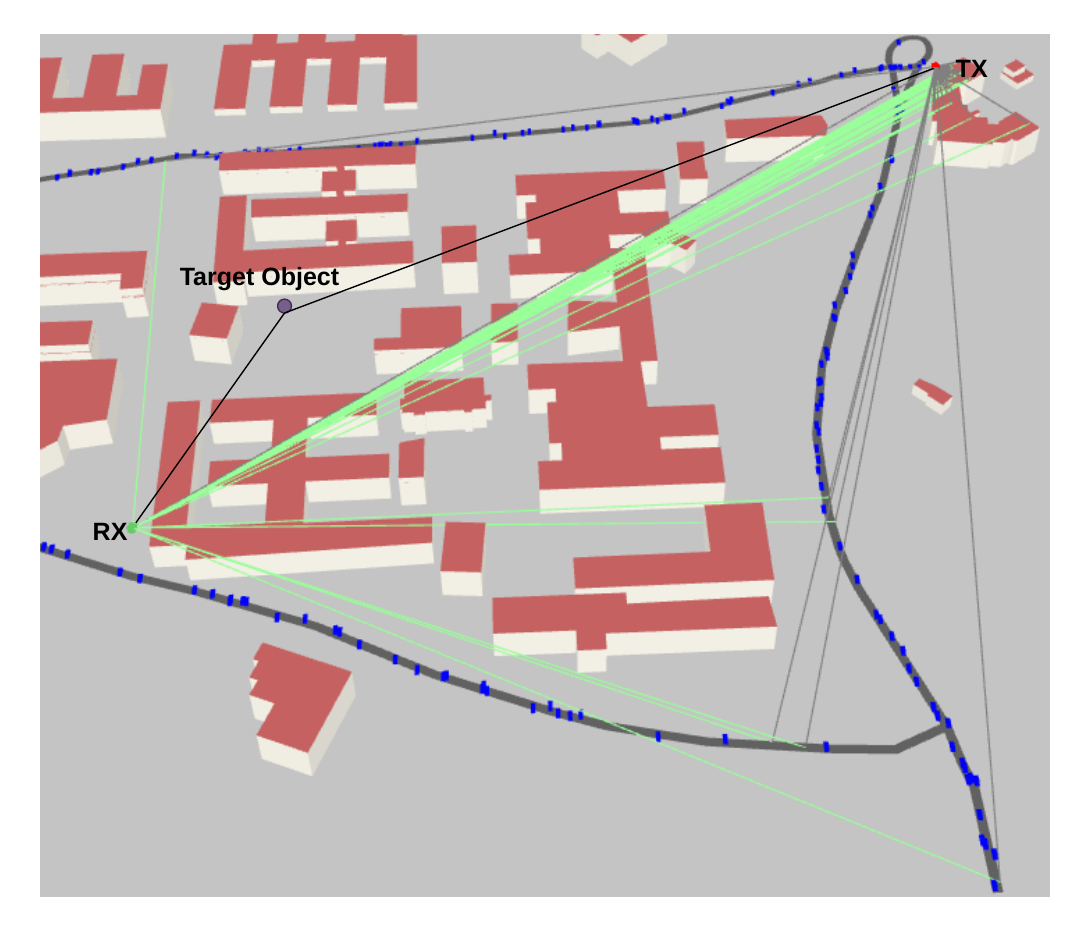}}
\caption{Cross-scene validation environment (Scene-IITM):
(a) IIT Madras campus layout imported from OpenStreetMap
(\textcopyright\ OpenStreetMap contributors, available under the Open
Database License~\cite{openstreetmap}); (b) TX--RX deployment with
$500$\,m separation.}
    \label{fig:IITM_all}
\end{figure}

\begin{figure*}[!t]
    \centering
    \includegraphics[
      width=0.85\textwidth,
      keepaspectratio,
      trim={0 0 0 0},
      clip]{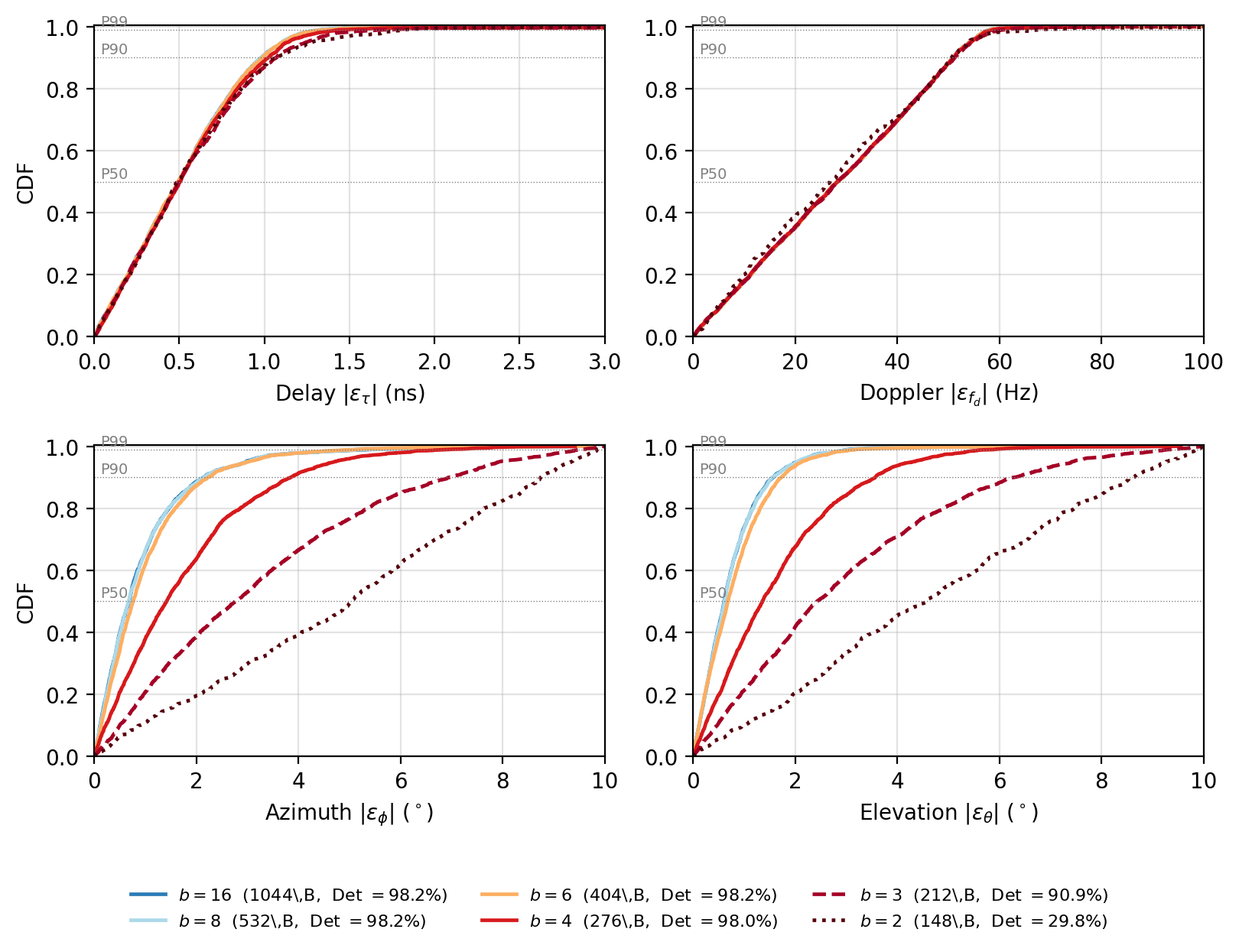}
    \caption{Per-link absolute-error CDFs at the STD operating point $(K,C){=}(8,64)$ on Scene-Synth, swept over post-training bit width $b\in\{16,8,6,4,3,2\}$. Delay and Doppler CDFs are quantization-invariant across the entire range, while angle errors are unaffected for $b\ge 6$, exhibit P90/P99 inflation at $b{=}4$, and collapse at $b\le 3$.}
    \label{fig:cdf_bsweep_std}
\end{figure*}

\begin{figure*}[!t]
    \centering
    \includegraphics[width=0.85\textwidth, keepaspectratio,
      trim={0 0 0 0}, clip]{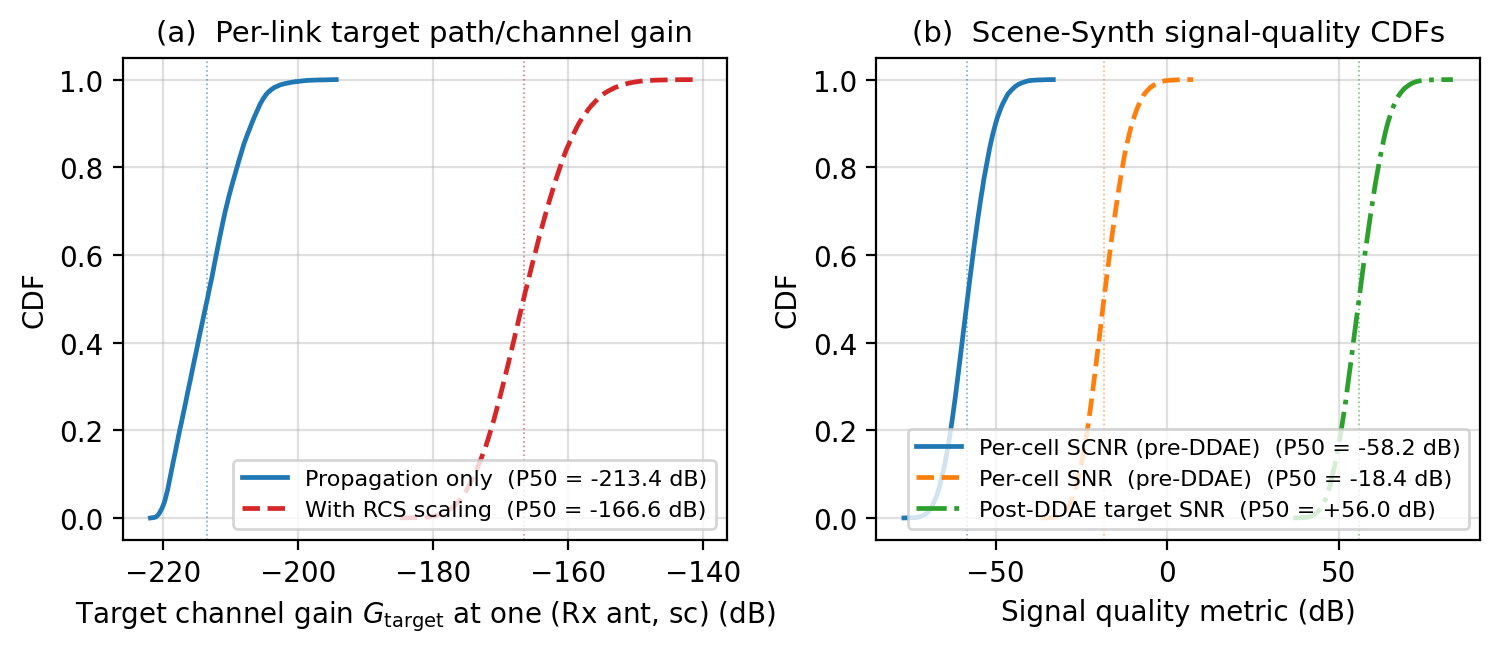}
\caption{Scene-Synth signal quality (24{,}000 link-instances).
(a)~Per-cell target channel-gain CDF, with/without RCS scaling.
(b)~Per-cell SCNR and SNR (pre-DDAE) and post-DDAE target SNR
(after coherent integration over Rx array, subcarriers, and symbols).
The SNR-to-post-DDAE gap ($\approx 74$~dB $= 10\log_{10}(N_r N_{sc} L)$)
is the coherent integration gain; the extra gain over the lower SCNR
reflects clutter suppression from delay--Doppler--angle focusing.}
    \label{fig:signal_quality_summary}
\end{figure*}

\begin{figure*}[!t]
    \centering
    \subfloat[\label{fig:dd_spectrum}]{%
        \includegraphics[width=0.45\textwidth,trim={0 0 0 0},clip]{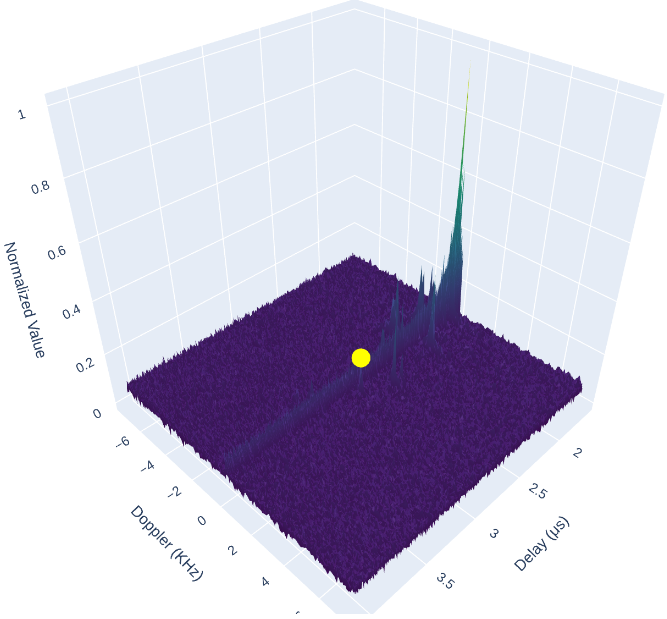}}
    \hfill
    \subfloat[\label{fig:ae_spectrum}]{%
        \includegraphics[width=0.45\textwidth, trim={0 30 0 0},clip]{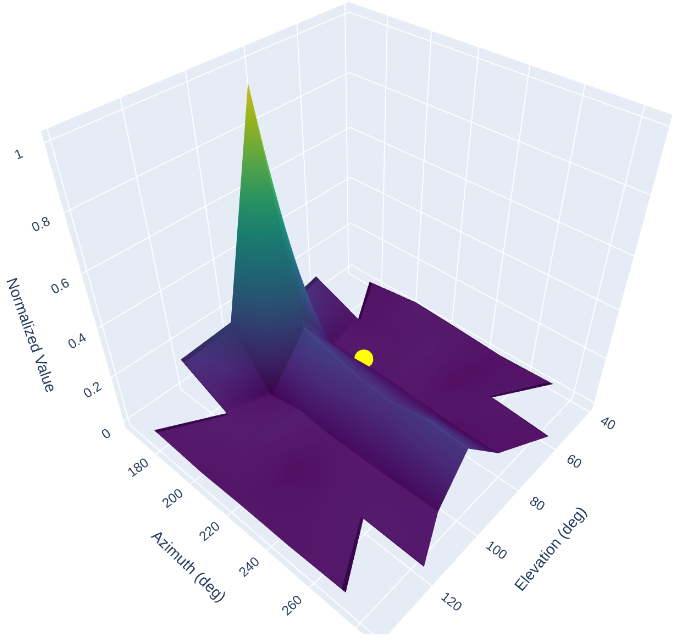}}
\caption{Representative peak-normalized spectra from a Scene-Synth sample: (a)~delay--Doppler spectrum; (b)~azimuth--elevation spectrum at the target delay--Doppler cell. The yellow marker indicates the discretized ground-truth target bin on each grid.}
    \label{fig:power_spectrum}
\end{figure*}

% \section{Experimental Setup}

\subsection{Synthetic-scene deployment and dataset generation}
\label{sec:scene_synth_setup}

For evaluation, a synthetic urban scene of size
$1000\,\mathrm{m}\times 1000\,\mathrm{m}$ is created in Blender,
as illustrated in Fig.~5. The environment contains roads,
cars, and building blocks, with building heights up to
$15\,\mathrm{m}$. Both roads and buildings are assigned ITU
concrete material properties. In addition, each road is modeled
as a 6-lane road with 3 lanes in each direction, and hundreds
of vehicles are randomly placed over the road network. Each
vehicle is modeled as an object of size
$[x_c,y_c,z_c]$ and assigned ITU-metal material properties.
Their positions are sampled randomly along the road lanes, and
their velocities are drawn randomly in the range
$[0,40]\,\mathrm{m/s}$. This scene is used to generate
site-specific sensing data using ray tracing, so that the
resulting channel parameters reflect deterministic reflections,
urban clutter, and geometry-dependent propagation effects
instead of purely statistical sampling. The underlying
propagation model follows 3GPP TR~38.901, while the
actual path parameters are obtained from NVIDIA Sionna
ray tracing.

The sensing deployment is based on a rectangular layout with
inter-site distance $500\,\mathrm{m}$. In the first bistatic
configuration shown in Fig.~5(a), the transmitter is placed at
$(250,250)\,\mathrm{m}$ and the receiver is placed at
$(250,-250)\,\mathrm{m}$. To provide additional geometric
diversity, a second transmitter location at
$(-250,-250)\,\mathrm{m}$ and a second receiver location at
$(-250,250)\,\mathrm{m}$ are also used, leading to the four
bistatic layouts shown in Fig.~5(a)--(d). TX and RX heights
are fixed at $25\,\mathrm{m}$. Across these layouts, the same
urban scene is observed from different bistatic perspectives,
which helps expose the model to different target signatures
and propagation conditions rather than allowing it to overfit
to a single fixed link geometry.

In the horizontal plane, the target coordinates are sampled uniformly in $[-240, 240]\,\mathrm{m}$ along both axes, rather than over the full $[-250, 250]\,\mathrm{m}$ extent, so that a small standoff is maintained from the TX/RX node locations. Along the vertical dimension, the target height is sampled in the range $[20, 300]\,\mathrm{m}$. To ensure that every valid target's bistatic delay falls within the retained processing window (Sec.~\ref{sec:sysmodel}, with $D = 1200$ bins corresponding to bistatic ranges in $[488, 1221]\,\mathrm{m}$), the sampling is additionally constrained to
\begin{equation}
\|\mathbf{p}_o - \mathbf{p}_t\| + \|\mathbf{p}_r - \mathbf{p}_o\| \leq 1220\,\mathrm{m}.
\label{eq:bistatic_range_limit}
\end{equation}
Candidate target positions exceeding this bistatic-range limit are rejected during dataset construction. This combined sampling region is consistent with the practical far-field assumption adopted in the system model and avoids near-collocated TX/RX-target cases. For each valid
TX--target--RX geometry, the corresponding target and
background channel parameters are computed using Sionna
ray tracing over the same urban layout, and the resulting
samples are used to form the synthetic dataset for training,
validation, and testing.

\subsection{Common simulation, waveform, and preprocessing settings}
\label{sec:common_waveform_setup}

For each selected TX--RX geometry, the target and background
channel parameters are generated using NVIDIA Sionna ray
tracing under the 3GPP TR~38.901 propagation framework.
The resulting channel impulse response is mapped to an
NR/OFDM channel frequency response (CFR), and sensing
snapshots are obtained from pilot-based least-squares channel
estimation on PRS resources. The estimated CFR snapshots are
then transformed into the DDAE representation described in
Sec.~II and converted to a quantized input tensor for the
learning-based estimator.

We use carrier frequency of $13\,\mathrm{GHz}$, subcarrier spacing
$120\,\mathrm{kHz}$, and an overall bandwidth of approximately
$400\,\mathrm{MHz}$, corresponding to $N_{sc}=3276$ active
subcarriers. One PRS OFDM symbol is used every $8$ OFDM
symbols. The coherent processing interval spans
$L_{\max}=1024$ OFDM symbols, which yields $L=128$
sensing snapshots for Doppler processing. The receiver employs
an $8\times 8$ UPA with half-wavelength spacing, while the
transmitter uses single-port transmission. After DDAE formation, we retain $D = 1200$ delay bins starting at offset $d_0 = 800$ ($\approx 1.63\,\mu\mathrm{s}$) and $V = 128$ Doppler bins centered at $v_0 = 64$. The retained delay window corresponds to $[1.63, 4.07]\,\mu\mathrm{s}$, or equivalently bistatic ranges of $[488, 1221]\,\mathrm{m}$, covering all bistatic delays produced by the target sampling region of Sec.~\ref{sec:setup}-A under the bistatic-range constraint enforced there; no geometrically-valid target is cut off at either end of the retained window. The lower edge is placed just below the TX--RX LOS arrival ($\approx 1.67\,\mu\mathrm{s}$) to allow for sidelobe leakage near LOS. The final network input has size $A \times E \times V \times D = 8 \times 8 \times 128 \times 1200$ and is represented using 8-bit uniform quantization. The common waveform and preprocessing parameters are summarized in Table~\ref{tab:common_setup}.

\begin{table}[t]
\centering
\caption{Common waveform and preprocessing parameters used in all experiments.}
\label{tab:common_setup}
\renewcommand{\arraystretch}{1.1}
\setlength{\tabcolsep}{5pt}
\begin{tabular}{p{5.0cm} p{2.3cm}}
\hline
\textbf{Parameter} & \textbf{Value} \\
\hline
Carrier frequency $f_c$ & 13 GHz \\
Subcarrier spacing (SCS) & 120 kHz \\
Active subcarriers $N_{sc}$ & 3276 \\
Delay FFT/grid length $N_{\tau}$ & 4096\\
Overall bandwidth & $\approx 400$ MHz \\
OFDM symbols in CPI $L_{\max}$ & 1024 ($\approx
8.53$ ms)\\
PRS snapshot spacing & 8 OFDM symbols \\
PRS snapshots $L$ & 128 \\
RX array size & $8 \times 8$ UPA \\
TX transmission & single-port \\
Retained delay bins $D$ & 1200 \\
Retained Doppler bins $V$ & 128 \\
Delay-bin offset $d_0$ & 800 \\
Doppler-bin center $v_0$ & 64 \\
Input tensor size $A \times E \times V \times D$ & $8 \times 8 \times 128 \times 1200$ \\
Input quantization & 8-bit \\
Delay-bin resolution $\Delta_{\text{del}}$    & $\approx 2.0345$~ns/bin \\
Doppler-bin resolution $\Delta_{\text{dop}}$  & $\approx 117.2$~Hz/bin \\
\hline
\end{tabular}
\end{table}

Unlike a link-level study with a fixed nominal operating SNR,
the present dataset does not admit a single representative SNR
value for the entire scene. For each TX--target--RX realization,
the received target SNR depends on the specific bistatic
geometry, propagation loss, ray-traced multipath/small-scale
fading, and the randomly drawn target RCS. Since target
locations are sampled over the deployment area and target RCS
is generated from the assumed statistical model, the received
target signal quality varies substantially across samples. Therefore, rather than reporting a single SNR number,
Fig.~\ref{fig:signal_quality_summary}(a) shows the empirical
CDF of the per-link target channel gain at a single
$(\text{Rx antenna}, \text{subcarrier})$ cell, computed from
the bistatic ray-tracing coefficients with and without RCS
scaling, and Fig.~\ref{fig:signal_quality_summary}(b) shows the
corresponding per-cell SCNR, per-cell SNR, and post-DDAE target
SNR after coherent integration over Rx antennas, subcarriers,
and OFDM symbols.
The post-DDAE target SNR has median (P50) and 90th-percentile
(P90) values of $55.96$ dB and $64.65$ dB, respectively,
summarizing the typical and upper-range operating conditions
induced by the considered deployment, propagation, and random
target-RCS sampling.

\subsection{Main model and training configuration}
\label{sec:main_model_training}

The trained model is evaluated at three named operating points
that span a $7.5\times$ payload range: \textbf{UC}
$(K, C, b){=}(4, 32, 6)$ at $107$~B per sample, \textbf{STD}
$(K, C, b){=}(8, 64, 6)$ at $404$~B, and \textbf{HF}
$(K, C, b){=}(16, 64, 6)$ at $806$~B. These anchors are
selected from a wider $(K, C, b)$ sweep over
$K \in \{2, 4, 8, 16, 32, 64, 128, 256, 512\}$,
$C \in \{4, 8, 16, 32, 64, 128\}$, and
$b \in \{32, 16, 8, 6, 4, 3, 2\}$. All anchors and the wider sweep
use the same model architecture. The power$+$peakedness DD
descriptor, the three-encoder candidate token
($g_{\mathrm{AE}}$, $g_{\mathrm{pos}}$, $g_{\mathrm{sc}}$), the
single-layer Set Transformer, the candidate reranker, and the
bounded sub-bin offset head described in
Sec.~\ref{sec:method} with the DD-context encoder removed
on the basis of the ablation study of
Sec.~\ref{sec:results}-\!\!~\ref{sec:ablation_studies}.

The dataset is split into $80\%$ training, $10\%$ validation,
and $10\%$ test subsets.
Training is performed for $25$ epochs with batch size $64$
using the AdamW optimizer, initial learning rate
$10^{-3}$, and weight decay $10^{-3}$. The dropout rate is set to 0.1 in the candidate-fusion MLP, the Set Transformer (attention weights and FFN output), the reranking MLP, and the angle-estimation head. During training, the reranking loss weight is
linearly warmed up over the first $5$ epochs to its full value.
Soft teacher forcing is scheduled from $1.0$ at the start of
training to $0.0$ at the final epoch, with mixing coefficient
$\alpha_{\mathrm{TF}}=0.9$. The reported
results correspond to the checkpoint selected by the best
validation metric.

\subsection{Evaluation protocol and metrics}
\label{sec:eval_metrics}

All results are reported on the held-out Scene-Synth test split.
Detection performance is summarized by detection rate and miss
rate. Estimation accuracy is measured using absolute error
$\varepsilon$ and mean absolute error (MAE). For delay and
Doppler,
\begin{equation}
\varepsilon_\tau = |\tilde \tau-\tau|,
\qquad
\varepsilon_{fd} = |\tilde f_d-f_d|,
\end{equation}
with delay reported in ns and Doppler in Hz. For azimuth
$\phi$ and elevation $\theta$, wrap-aware angular errors are
used:
\begin{equation}
\varepsilon_{\phi}
=
|\operatorname{wrap}_{[-\pi,\pi]}(\tilde\phi-\phi)|,
\qquad
\varepsilon_{\theta}
=
|\operatorname{wrap}_{[-\pi,\pi]}(\tilde\theta-\theta)|,
\end{equation}
and the corresponding MAEs are reported in degrees.

A test sample is declared detected if all four estimation
errors satisfy fixed thresholds,
\[
\varepsilon_\tau \le \tau_{c},\qquad
\varepsilon_{fd} \le f_{dc},\qquad
\varepsilon_{\phi} \le \phi_c,\qquad
\varepsilon_{\theta} \le \theta_c,
\]
where $(\tau_c,f_{dc},\phi_c,\theta_c)
= (50 \text{ ns}, 200 \text{ Hz}, 10^{\circ}, 10^{\circ})$
are the detection thresholds used in evaluation. Detection rate
and miss rate are computed over the full test set using this rule,
whereas estimation errors are reported over detected
link-instances only.

Error distributions are summarized using percentile statistics.
For an absolute error variable $e$, the median error $P50$ is
the 50th percentile, and $P90$ satisfies
\[
\Pr(e \le P90)=0.9,
\]
i.e., $90\%$ of the evaluated samples have error not exceeding
$P90$. In the main, operating-point, and ablation tables, we
report MAE and $P90$; for the quantized-feedback results, we
additionally report $P50$. For compact summary plots and
selected operating-point comparisons, we also use the
angle-sum metric $\mathrm{MAE}_{\phi}+\mathrm{MAE}_{\theta}$.

\subsection{Baselines and Quantized-Feedback Evaluation}
We compare the proposed learned estimator against a classical peak-selection baseline, including variants with and without static-clutter suppression. Static-clutter removal is important for the classical baseline, but even the filtered version remains substantially weaker than the learned approach in detection-oriented performance. We therefore use the filtered classical baseline as the main classical reference in the paper.

For feedback-compression experiments, quantization is applied \emph{after training} and \emph{during inference} on the trained full-precision model. Specifically, latent quantization is inserted after candidate token construction and before reranking and aggregation at the SF, so that the receiver-side inference pipeline is unchanged except that it operates on reconstructed latents. The implementation uses a \emph{global per-sample scale} by default, i.e., one $B_{\mathrm{side}}$ for the full $K\times C$ latent matrix.

Payload sweeps at fixed $K$ and $C$ are reported using the \emph{latent payload} $KCb$. When the full feedback payload is needed, including the candidate-index overhead and one global scale parameter $B_{\mathrm{side}}$,
\begin{equation*}
B_{\mathrm{fb}} = KCb + 18K + B_{\mathrm{side}},
\end{equation*}
because the delay--Doppler grid has size $V=128$ and $D=1200$, so each candidate index requires
\begin{equation*}
\left\lceil \log_2(VD) \right\rceil
=
\left\lceil \log_2(128 \times 1200) \right\rceil
= 18
\end{equation*}
bits. With one $16$-bit global scale ($B_{\mathrm{side}}=16$), the STD anchor $(K, C, b){=}(8, 64, 6)$ has latent payload $3{,}072$~bits, index overhead $144$~bits, and scale $16$~bits, totaling $3{,}232$~bits or $404$~bytes per sample. The same expression evaluated at UC and HF gives $107$~B and $806$~B per sample, respectively.

This convention is important when interpreting the quantized-feedback tables and payload--accuracy plots in Sec.~\ref{sec:quant_sensitivity}: when $K$ and $C$ are fixed, showing latent payload alone is acceptable, provided the caption clearly states that index and scale overheads are constant and omitted.

\section{Results}
\label{sec:results}

This section evaluates the proposed coarse-to-fine candidate-latent codec
on the Scene-Synth held-out test split and reports four complementary
findings: (i) the codec reduces the per-link SE--SF feedback to
107--806 bytes per CPI across the three named anchors, corresponding to
sub-Mbit/s interface rates and large reductions relative to raw CFR or
DDAE-magnitude forwarding; (ii) the resulting accuracy is essentially
insensitive to post-training latent quantization down to $b=6$ and
degrades gracefully toward $b=4$ subject to a latent-dimension condition;
(iii) the trained model transfers to an independent campus-scale scene
(Scene-IITM) without fine-tuning; and (iv) each architectural component
contributes a measurable share of the gain.

Throughout this section, three named operating points are
used as anchors:
\begin{itemize}
\item \textbf{UC} (ultra-compact): $(K,C,b)=(4,32,6)$, payload
$107$~B per sample, $\sim$$91{,}900\times$ compression
relative to the uncompressed 8-bit DDAE magnitude tensor;
\item \textbf{STD} (standard): $(K,C,b)=(8,64,6)$, payload
$404$~B per sample, $\sim$$24{,}300\times$ compression;
\item \textbf{HF} (high-fidelity): $(K,C,b)=(16,64,6)$, payload
$806$~B per sample, $\sim$$12{,}200\times$ compression.
\end{itemize}
The three anchors are intentionally placed across a $7.5\times$
payload range so that a downstream system designer can pick the
operating point most consistent with its fronthaul budget; the
paper does not commit to a single recommended point. The
selection of these anchors from a wider $(K,C,b)$ design space
is presented in Secs.~\ref{sec:design_studies} and
\ref{sec:quant_sensitivity}.

\subsection{Comparison with classical peak-selection baselines}
\label{sec:headline}

\begin{table*}[!tbp]
\centering
\caption{Comparison of the proposed coarse-to-fine candidate-latent codec against classical peak-selection baselines on Scene-Synth. The proposed scheme is reported at three named operating points (UC, STD, HF) spanning payload range, each at the quantized latent width $b{=}6$ on the no-DD-context architecture. Detection rate is over the full test set ($2{,}400$ link-instances, all satisfying $d_{t}+d_{r}\le 1220$\,m); estimation errors are MAE and $P90$ over detected link-instances. The classical-baseline payload entry of $16$\,B corresponds to the four estimated parameters $(\tau,\,f_d,\,\phi,\,\theta)$ transmitted as $32$-bit floats.}
\label{tab:baseline_static_removal}
\setlength{\tabcolsep}{3pt}
\renewcommand{\arraystretch}{1.15}
\resizebox{\textwidth}{!}{%
\begin{tabular}{l c c cc cc cc cc}
\toprule
\multirow{2}{*}{Method} &
\multirow{2}{*}{\makecell{Payload\\(B)}} &
\multirow{2}{*}{Det.\ (\%)} &
\multicolumn{2}{c}{Delay $\varepsilon_\tau$ (ns)} &
\multicolumn{2}{c}{Doppler $\varepsilon_{fd}$ (Hz)} &
\multicolumn{2}{c}{Azimuth $\varepsilon_\phi$ ($^\circ$)} &
\multicolumn{2}{c}{Elevation $\varepsilon_\theta$ ($^\circ$)} \\
\cmidrule(lr){4-5}\cmidrule(lr){6-7}\cmidrule(lr){8-9}\cmidrule(lr){10-11}
& & & MAE & P90 & MAE & P90 & MAE & P90 & MAE & P90 \\
\midrule
\multicolumn{11}{l}{\emph{Classical peak-selection baselines}} \\
Peak selection, no clutter removal           & 16 &  0.27 & 16.89 & 37.70 & 88.44 & 177.84 & 7.42 & 9.18 & 3.24 & 7.30 \\
Peak selection, with static removal          & 16 & 12.17 &  0.45 &  0.84 & 25.83 &  50.28 & 4.03 & 7.80 & 3.51 & 6.70 \\
\midrule
\multicolumn{11}{l}{\emph{Proposed candidate-latent, $b=6$}} \\
\textbf{UC}\quad $(K,C)=(4,32)$              &    107 & 96.33 & 0.54 & 0.98 & 28.47 & 50.70 & 0.90 & 1.92 & 0.77 & 1.64 \\
\textbf{STD}\quad $(K,C)=(8,64)$             &    404 & 98.17 & 0.57 & 0.99 & 28.54 & 50.88 & 1.05 & 2.24 & 0.84 & 1.73 \\
\textbf{HF}\quad $(K,C)=(16,64)$             &    806 & 98.88 & 0.53 & 0.93 & 28.56 & 50.79 & 1.05 & 2.15 & 0.77 & 1.61 \\
\bottomrule
\end{tabular}%
}
\end{table*}

Table~\ref{tab:baseline_static_removal} reports the three
named anchors alongside two classical peak-selection baselines
on Scene-Synth. Without static-clutter handling, peak selection
achieves only $0.27\%$ detection with delay MAE $16.89$~ns and
azimuth/elevation MAE $7.42^\circ$/$3.24^\circ$. Adding
static-clutter suppression lifts detection to $12.17\%$ at
delay MAE $0.45$~ns but leaves azimuth/elevation MAE above
$3.5^\circ$. By contrast, the three learned anchors reach
$96.33$/$98.17$/$98.88\%$ detection at delay MAE $\le 0.60$~ns
and azimuth/elevation MAE below $1.2^\circ$. The Doppler results should be interpreted conditionally. The filtered peak-selection baseline gives a comparable Doppler MAE of $25.83$ Hz, but only over the small subset of samples it detects, with a joint detection rate of $12.17\%$. In contrast, the proposed UC, STD, and HF anchors maintain comparable Doppler MAE values of about $28.5$ Hz while increasing the joint detection rate to $96.33-98.88\%$.

Fig.~\ref{fig:power_spectrum} provides the qualitative
explanation for why classical peak selection cannot solve this
task. The delay--Doppler spectrum in
Fig.~\ref{fig:power_spectrum}(a) is normalized by its maximum
amplitude, and the azimuth--elevation spectrum in
Fig.~\ref{fig:power_spectrum}(b), taken at the target
delay--Doppler cell, is likewise normalized by its peak. The
yellow marker indicates the discretized ground-truth bin. In
this representative sample, the dominant spectral peak does
not coincide with the target location, and the target return
is not separable from clutter and interfering scatterers by
simple thresholding. Static-clutter suppression removes only
the dominant invariant features and is therefore necessary but
not sufficient. The learned codec resolves both issues by
reporting $K$ candidate hypotheses jointly with a low-dimensional
latent that summarizes their angular and contextual content,
rather than committing to a single peak.

\begin{table}[t]
\centering
\caption{Per-link SE--SF feedback payload and required interface throughput
for one CPI. The CPI duration is $T_{\rm CPI}=1024/120\,000\approx
8.53$ ms.}
\label{tab:feedback_payload_rate}
\setlength{\tabcolsep}{3.5pt}
\renewcommand{\arraystretch}{1.08}
\begin{tabular}{lcc}
\toprule
\textbf{Feedback representation} &
\textbf{Payload} &
\textbf{Required rate} \\
 & \textbf{(B/link/CPI)} & \textbf{(per link)} \\
\midrule
Raw complex CFR, 16-bit I + 16-bit Q
& $107{,}347{,}968$ & $100.6$ Gbit/s \\
DDAE magnitude, FP32 real
& $39{,}321{,}600$ & $36.86$ Gbit/s \\
DDAE magnitude, 8-bit real
& $9{,}830{,}400$ & $9.216$ Gbit/s \\
\midrule
UC: $(K,C,b)=(4,32,6)$
& $107$ & $0.100$ Mbit/s \\
STD: $(K,C,b)=(8,64,6)$
& $404$ & $0.379$ Mbit/s \\
HF: $(K,C,b)=(16,64,6)$
& $806$ & $0.756$ Mbit/s \\
\bottomrule
\end{tabular}
\end{table}

At broadside, an $8\times 8$ UPA has angular bin width $\approx 14.3^\circ$, giving a peak-only readout floor of $W/4 \approx 3.58^\circ$ under perfect bin selection. The proposed angular encoder feeds a differentiable spatial-softmax estimate of the sub-bin peak location into the candidate latent, which the SF-side head decodes as $(\sin,\cos)$ of the continuous angle. At the STD anchor this yields azimuth and elevation MAEs of $1.05^\circ$ and $0.84^\circ$ --- roughly $3.4\times$ and $4.2\times$ below the broadside floor. Off-broadside bins are wider by $1/\cos\theta$, so the effective super-resolution factor over the target distribution is larger than these broadside values.

Table~\ref{tab:feedback_payload_rate} places the proposed
candidate-latent feedback in the context of two direct-forwarding
references. Forwarding the raw complex CFR over one CPI requires
$8\times8\times3276\times128$ complex samples per TX--RX link.
With 16-bit in-phase and 16-bit quadrature components, this amounts
to $107.35$ MB per link per CPI, or about $100.6$ Gbit/s if the
feedback must be delivered within the CPI. Even forwarding only the
DDAE magnitude remains expensive: the retained
$8\times8\times128\times1200$ DDAE grid requires $39.32$ MB as
FP32 real values, or $9.83$ MB as 8-bit magnitudes, corresponding to
$36.86$ Gbit/s and $9.216$ Gbit/s, respectively. In contrast, the
UC, STD, and HF candidate-latent anchors require only 107, 404, and
806 bytes per link per CPI, reducing the required SE--SF throughput
to sub-Mbit/s levels while preserving the detection and estimation
performance reported above.

\subsection{Operating-point sweep at FP32 latents}
\label{sec:design_studies}

The proposed pipeline exposes three design variables that
together determine the SE--SF feedback payload: the candidate
count $K$, the latent dimension $C$, and the latent
quantization bit width $b$. Their joint effect is captured by
the bit-budget expression of \eqref{eq:budget_general}, which
for the considered delay--Doppler grid reduces to
$B_{\mathrm{fb}} = bKC + 18K + 16$~bits. Anchor selection in
this paper proceeds in two phases: a wide $(K,C)$ sweep at
FP32 latents in this subsection, followed by a quantization
sweep over $b$ on the resulting shortlist in
Sec.~\ref{sec:quant_sensitivity}.

\begin{table}[!t]
\centering
\caption{Operating-point sweep over $K$ and $C$ on Scene-Synth at FP32 latents.}
\label{tab:kc_full_fp32}
\setlength{\tabcolsep}{4pt}
\renewcommand{\arraystretch}{1.05}
\footnotesize
\begin{tabular}{r r r r r r r r r}
\toprule
$K$ & $C$ & $KC$ & \makecell[r]{Pay.\\(B)} & \makecell[r]{Det\\(\%)} & \makecell[r]{$\varepsilon_\tau$\\(ns)} & \makecell[r]{$\varepsilon_{fd}$\\(Hz)} & \makecell[r]{$\varepsilon_\phi$\\($^\circ$)} & \makecell[r]{$\varepsilon_\theta$\\($^\circ$)} \\
\midrule
   2 &   16 &   32 &   134 & 92.83 & 0.5 & 29.1 & 0.95 & 1.01 \\
   2 &   32 &   64 &   262 & 92.25 & 0.5 & 28.8 & 0.85 & 0.59 \\
   2 &   64 &  128 &   518 & 93.21 & 0.5 & 28.8 & 0.94 & 0.63 \\
   2 &  128 &  256 &  1030 & 91.42 & 0.5 & 29.0 & 1.12 & 0.93 \\
\midrule
   4 &    8 &   32 &   139 & 95.00 & 0.5 & 28.9 & 1.17 & 0.92 \\
   4 &   16 &   64 &   267 & 95.88 & 0.6 & 29.1 & 0.95 & 0.97 \\
   4 &   32 &  128 &   523 & 96.25 & 0.5 & 28.5 & 0.80 & 0.67 \\
   4 &   64 &  256 &  1035 & 96.04 & 0.6 & 28.8 & 0.82 & 0.73 \\
   4 &  128 &  512 &  2059 & 95.83 & 0.5 & 28.9 & 1.16 & 0.94 \\
\midrule
   8 &    4 &   32 &   148 & 96.83 & 0.7 & 28.8 & 1.53 & 1.38 \\
   8 &    8 &   64 &   276 & 97.04 & 0.7 & 28.6 & 1.14 & 1.20 \\
   8 &   16 &  128 &   532 & 97.21 & 0.6 & 28.7 & 0.87 & 0.92 \\
   8 &   32 &  256 &  1044 & 97.71 & 0.6 & 28.6 & 1.19 & 1.01 \\
   8 &   64 &  512 &  2068 & 98.21 & 0.5 & 28.5 & 0.98 & 0.77 \\
   8 &  128 & 1024 &  4116 & 97.58 & 0.5 & 28.6 & 1.00 & 0.91 \\
\midrule
  16 &    4 &   64 &   294 & 97.25 & 0.8 & 28.6 & 1.52 & 1.54 \\
  16 &    8 &  128 &   550 & 98.71 & 0.6 & 28.4 & 1.22 & 0.98 \\
  16 &   16 &  256 &  1062 & 98.54 & 0.5 & 28.7 & 0.99 & 0.91 \\
  16 &   32 &  512 &  2086 & 98.25 & 0.5 & 28.5 & 1.17 & 1.13 \\
  16 &   64 & 1024 &  4134 & 98.92 & 0.5 & 28.6 & 0.98 & 0.69 \\
\midrule
  32 &    4 &  128 &   586 & 97.29 & 0.6 & 28.6 & 1.36 & 1.20 \\
  32 &    8 &  256 &  1098 & 97.67 & 0.6 & 28.7 & 1.15 & 0.99 \\
  32 &   16 &  512 &  2122 & 98.62 & 0.5 & 28.4 & 0.98 & 1.17 \\
  32 &   32 & 1024 &  4170 & 98.38 & 0.6 & 28.6 & 1.23 & 1.03 \\
  32 &   64 & 2048 &  8266 & 98.38 & 0.6 & 28.4 & 0.87 & 0.74 \\
\midrule
  64 &    4 &  256 &  1170 & 97.58 & 0.6 & 28.3 & 1.54 & 1.40 \\
  64 &    8 &  512 &  2194 & 98.58 & 0.6 & 28.6 & 1.25 & 0.89 \\
  64 &   16 & 1024 &  4242 & 98.75 & 0.5 & 28.6 & 1.10 & 0.94 \\
  64 &   32 & 2048 &  8338 & 99.21 & 0.5 & 28.6 & 1.04 & 0.75 \\
\midrule
 128 &    4 &  512 &  2338 & 97.62 & 0.7 & 28.4 & 1.73 & 1.33 \\
 128 &    8 & 1024 &  4386 & 98.50 & 0.6 & 28.0 & 1.30 & 1.10 \\
 128 &   16 & 2048 &  8482 & 98.79 & 0.6 & 28.7 & 0.97 & 1.12 \\
\midrule
 256 &    4 & 1024 &  4674 & 96.42 & 0.6 & 28.1 & 1.69 & 1.67 \\
 256 &    8 & 2048 &  8770 & 98.83 & 0.6 & 28.3 & 1.14 & 1.05 \\
 256 &   16 & 4096 & 16962 & 98.54 & 0.6 & 28.4 & 1.15 & 1.02 \\
\midrule
 512 &    4 & 2048 &  9346 & 97.21 & 0.6 & 28.2 & 1.68 & 1.70 \\
 512 &    8 & 4096 & 17538 & 99.17 & 0.6 & 28.4 & 1.33 & 1.00 \\
\bottomrule
\end{tabular}
\end{table}

Table~\ref{tab:kc_full_fp32} reports end-to-end performance
over a $(K,C)$ grid spanning
$K \in \{2, 4, 8, 16, 32, 64, 128, 256, 512\}$ and
$C \in \{4, 8, 16, 32, 64, 128\}$, for a total of $37$
trained configurations under the no-DD-context architecture
adopted in Sec.~\ref{sec:ablation_studies}. All entries are
evaluated with FP32 latents and no post-training quantization,
so the payload column reflects the latent contribution
$32KC$~bits plus the constant overhead $18K+16$~bits.

\emph{(i) Candidate-count ceiling.} The detection rate is upper
bounded by $K$ alone, independent of $C$. At $K=2$ the model
ceilings near $92\%$ across all evaluated $C$, at $K=4$ near
$95.5\%$, at $K=8$ near $97.5\%$. From $K=16$ onward the
detection rate stabilizes around $98$--$98.7\%$, and additional
candidates yield diminishing returns at increasing
candidate-index cost.

\emph{(ii) Latent-dimension controls angular accuracy.} For
fixed $K \ge 8$, azimuth and elevation MAEs improve as $C$
grows, while delay and Doppler MAEs are nearly
$C$-invariant. A per-link delay MAE of
$0.5$--$0.7$~ns and a Doppler MAE near $28.5$~Hz appear at
essentially every grid point, set by the sub-bin estimator and
the OFDM observation window, respectively. Angular accuracy,
by contrast, depends on the codec's ability to encode antenna-aperture
structure, which is gated by $C$. The cleanest evidence is the
$(K{=}16, C)$ row of Table~\ref{tab:kc_full_fp32}: azimuth MAE
drops from $1.52^\circ$ at $C=4$ to $0.98^\circ$ at $C=64$
while delay MAE stays within $0.3$~ns.

\emph{(iii) Index overhead at small $C$.} The candidate-index
contribution $18K$ to the payload is linear in $K$ and
independent of $C$. At $C=4$ this overhead is comparable to
or larger than the latent budget, so doubling $C$ from $4$ to
$8$ doubles the latent capacity at much smaller payload cost
than doubling $K$. The pair
$(K, C){=}(16, 32)$ at $2{,}086$~B FP32 and
$(K, C){=}(32, 16)$ at $2{,}122$~B FP32 illustrate this:
both encode the same $KC=512$ latent footprint but with very
different latent/index ratios; both reach $\sim$$98\%$
detection on Scene-Synth, confirming that for a fixed $KC$
budget smaller $K$ (larger $C$) is preferable.

Tracing the (payload, detection) trade-off curve in
Table~\ref{tab:kc_full_fp32}, we select seven $(K, C)$ anchors
spanning the relevant operating range:
$(4, 16)$, $(4, 32)$, $(8, 32)$, $(8, 64)$, $(16, 32)$, $(16, 64)$,
and $(64, 32)$. These cover candidate counts from $K = 4$ to
$K = 64$ and latent dimensions from $C = 16$ to $C = 64$, with
FP32 payloads ranging from $267$~B at $(4, 16)$ to $8{,}338$~B at
$(64, 32)$ and detection rates from $95.88\%$ to $99.21\%$. This
shortlist is carried into the quantization study of
Sec.~\ref{sec:quant_sensitivity}.

\subsection{Quantization sensitivity}
\label{sec:quant_sensitivity}

\begin{table*}[!tbp]
\centering
\caption{Per-parameter estimation statistics at the seven shortlisted $(K,C)$ anchors on Scene-Synth, reported at the named operating point ($b{=}6$) and under the aggressive quantization stress test ($b{=}4$). Payload is the per-sample feedback budget $B_{\mathrm{fb}}=bKC+18K+16$ bits. Detection rate is computed over the full test set ($2{,}400$ link-instances); per-parameter errors are MAE and 50/90/99th percentiles over detected link-instances only.}
\label{tab:kcb_anchors_full}
\setlength{\tabcolsep}{2.5pt}
\renewcommand{\arraystretch}{1.1}
\resizebox{\textwidth}{!}{%
\begin{tabular}{r r r r | r r r r | r r r r | r r r r | r r r r}
\toprule
\multirow{2}{*}{$K$} & \multirow{2}{*}{$C$} &
\multirow{2}{*}{\makecell{Pay.\\(B)}} & \multirow{2}{*}{\makecell{Det\\(\%)}} &
\multicolumn{4}{c}{Delay $\varepsilon_\tau$ (ns)} &
\multicolumn{4}{c}{Doppler $\varepsilon_{fd}$ (Hz)} &
\multicolumn{4}{c}{Azimuth $\varepsilon_\phi$ ($^\circ$)} &
\multicolumn{4}{c}{Elevation $\varepsilon_\theta$ ($^\circ$)} \\
\cmidrule(lr){5-8}\cmidrule(lr){9-12}\cmidrule(lr){13-16}\cmidrule(lr){17-20}
& & & & MAE & P50 & P90 & P99 & MAE & P50 & P90 & P99 & MAE & P50 & P90 & P99 & MAE & P50 & P90 & P99 \\
\midrule
\multicolumn{20}{l}{\emph{Named operating point ($b=6$)}} \\
\midrule
   4 &  16 &    59 & 95.83 & 0.55 & 0.50 & 1.00 & 1.37 & 29.12 & 29.05 & 51.47 & 60.62 & 1.18 & 0.86 & 2.49 & 5.29 & 1.18 & 0.95 & 2.42 & 4.71 \\
   4 &  32 &   107 & 96.33 & 0.54 & 0.51 & 0.98 & 1.48 & 28.47 & 28.32 & 50.70 & 58.97 & 0.90 & 0.65 & 1.92 & 5.21 & 0.77 & 0.61 & 1.64 & 2.97 \\
   8 &  32 &   212 & 97.75 & 0.61 & 0.49 & 0.96 & 1.73 & 28.65 & 28.29 & 51.08 & 59.48 & 1.26 & 1.01 & 2.54 & 5.66 & 1.08 & 0.80 & 2.28 & 5.29 \\
   8 &  64 &   404 & 98.17 & 0.57 & 0.49 & 0.99 & 1.36 & 28.54 & 28.22 & 50.88 & 58.20 & 1.05 & 0.77 & 2.24 & 5.15 & 0.84 & 0.68 & 1.73 & 3.20 \\
  16 &  32 &   422 & 98.29 & 0.54 & 0.50 & 0.95 & 1.21 & 28.55 & 28.38 & 50.98 & 58.93 & 1.25 & 0.95 & 2.60 & 5.90 & 1.19 & 0.88 & 2.56 & 5.52 \\
  16 &  64 &   806 & 98.88 & 0.53 & 0.51 & 0.93 & 1.28 & 28.56 & 28.46 & 50.79 & 58.74 & 1.05 & 0.82 & 2.15 & 4.78 & 0.77 & 0.59 & 1.61 & 3.31 \\
  64 &  32 &  1682 & 99.21 & 0.52 & 0.50 & 0.92 & 1.16 & 28.54 & 28.59 & 51.16 & 57.90 & 1.17 & 0.94 & 2.33 & 5.07 & 0.92 & 0.73 & 1.88 & 4.40 \\
\midrule
\multicolumn{20}{l}{\emph{Quantization stress test ($b=4$)}} \\
\midrule
   4 &  16 &    43 & 93.75 & 0.56 & 0.50 & 1.11 & 1.51 & 29.12 & 28.67 & 52.23 & 65.13 & 2.58 & 2.15 & 5.46 & 8.50 & 2.66 & 2.18 & 5.55 & 9.11 \\
   4 &  32 &    75 & 96.25 & 0.54 & 0.50 & 1.01 & 1.49 & 28.58 & 28.21 & 51.07 & 60.73 & 1.61 & 1.26 & 3.36 & 6.28 & 1.68 & 1.34 & 3.55 & 6.01 \\
   8 &  32 &   148 & 97.54 & 0.59 & 0.51 & 1.00 & 1.76 & 28.61 & 28.31 & 50.95 & 59.75 & 1.86 & 1.44 & 3.98 & 7.38 & 1.65 & 1.28 & 3.53 & 6.51 \\
   8 &  64 &   276 & 98.04 & 0.59 & 0.51 & 1.02 & 1.44 & 28.50 & 28.41 & 51.02 & 58.73 & 1.79 & 1.42 & 3.85 & 6.97 & 1.67 & 1.36 & 3.57 & 5.91 \\
  16 &  32 &   294 & 98.17 & 0.57 & 0.51 & 0.94 & 1.29 & 28.45 & 28.04 & 50.87 & 59.10 & 1.91 & 1.49 & 4.14 & 7.43 & 1.79 & 1.42 & 3.84 & 7.06 \\
  16 &  64 &   550 & 98.62 & 0.55 & 0.51 & 0.95 & 1.39 & 28.45 & 28.21 & 51.12 & 58.81 & 1.83 & 1.49 & 3.81 & 6.90 & 1.46 & 1.16 & 3.14 & 5.40 \\
  64 &  32 &  1170 & 98.29 & 0.52 & 0.49 & 0.93 & 1.20 & 28.56 & 28.38 & 51.25 & 58.73 & 2.25 & 1.86 & 4.71 & 7.89 & 2.32 & 1.82 & 5.05 & 8.44 \\
\bottomrule
\end{tabular}%
}
\end{table*}

Post-training quantization is applied at inference time on the
trained FP32 model using a uniform mid-rise quantizer with one
global per-sample scale, as specified in
Sec.~\ref{subsec:feedback}. No retraining or fine-tuning is
performed. Table~\ref{tab:kcb_anchors_full} reports full
per-parameter statistics (MAE, $P50$, $P90$, $P99$) at the
seven $(K, C)$ shortlist anchors of
Sec.~\ref{sec:design_studies}, at the named operating point
$b=6$ and at the aggressive stress point $b=4$.

\emph{Free-quantization regime ($b\ge 6$).} For every
shortlist anchor, detection at $b \in \{16, 8, 6\}$ is
statistically indistinguishable from FP32. At the STD anchor,
the detection rate is $98.17\%$ at $b=6$ versus $98.21\%$ at
FP32 (a $0.04$~pp difference),
while azimuth MAE rises from $0.98^\circ$ at FP32 to
$1.05^\circ$ at $b=6$ and elevation MAE from $0.77^\circ$ to
$0.84^\circ$. Across the shortlist, $b=8$ is within
$0.03$~pp of FP32 detection, and $b=6$ is within $0.10$~pp.
This is the practical statement of the operating-point choice:
$b=6$ extracts a $\sim$$5.3\times$ further payload reduction
over FP32 latents at essentially zero accuracy cost.

\textit{Stress regime} ($b = 4$). At $b = 4$ the angle tails
inflate visibly while delay and Doppler statistics remain
unchanged. For the STD anchor, the azimuth $P99$ rises from
$5.15^\circ$ at $b = 6$ to $6.97^\circ$ at $b = 4$ and the
elevation $P99$ from $3.20^\circ$ to $5.91^\circ$, while
detection stays at $98.04\%$. The same monotone tail inflation
appears at every shortlisted anchor in Table~\ref{tab:kcb_anchors_full},
and all seven anchors retain detection rates above $93.7\%$ at
$b = 4$, with the median and $P90$ angular errors remaining
below $2.2^\circ$ and $5.6^\circ$ respectively.

\emph{Cliff at $b \le 3$.} Below $b=4$, detection collapses
for every anchor. At the STD anchor, $b=3$ gives $90.9\%$
detection at $212$~B and $b=2$ gives $29.8\%$ at $148$~B
(Fig.~\ref{fig:cdf_bsweep_std}); conditional-on-detection
error metrics become unreliable in this regime. We therefore
report $b \le 3$ only for completeness and identify $b=4$ as
the lower envelope of the operating range.

Fig.~\ref{fig:cdf_bsweep_std} visualizes the full sweep
$b \in \{16, 8, 6, 4, 3, 2\}$ at the STD anchor as
per-parameter CDFs over the $2{,}400$ link instances. The
delay and Doppler CDFs are quantization-invariant across the
entire range, which makes the localization of quantization
noise in the angular channels explicit. The azimuth and
elevation CDFs cluster tightly for $b \in \{16, 8, 6\}$,
separate at $b=4$, and collapse at $b \le 3$; the inter-curve
gap is largest in the $P90$--$P99$ region.

\subsubsection*{Final anchor selection}
We select three named anchors from the shortlist of Table~\ref{tab:kcb_anchors_full}, each defined by a single criterion at $b{=}6$.
\textbf{UC} $(K,C,b){=}(4,32,6)$ at $107$~B is the smallest payload
at which the median angular error drops below $0.7^\circ$ while detection
remains above $95\%$; it dominates the smaller $(4,16,6)$ at $59$~B on
every angular statistic at a modest payload premium.
\textbf{STD} $(K,C,b){=}(8,64,6)$ at $404$~B is the smallest payload
reaching the $98\%+$ detection plateau.
\textbf{HF} $(K,C,b){=}(16,64,6)$ at $806$~B is the highest-detection
configuration within the sub-kilobyte envelope, reaching $98.88\%$
detection with the lowest azimuth $P99$ in the shortlist ($4.78^\circ$)
and a tight elevation $P99$ of $3.31^\circ$.

\subsection{Cross-scene generalization to Scene-IITM}
\label{sec:cross_scene_iitm}

To test whether the learned candidate-latent representation
generalizes beyond Scene-Synth, the trained model is evaluated
on Scene-IITM, an independent campus-scale scene constructed
from a 3-D IIT Madras layout (Fig.~\ref{fig:IITM_all}). The
scene uses the same material assumptions, waveform, PRS
structure, array configuration, preprocessing pipeline, and
the same trained model; only the building geometry, scatterer
field, and TX--RX placement (separation $500$~m, transmitter
near the Central Library and receiver on the EE Department
building) differ. No fine-tuning or scene-specific
re-training is performed. The test set contains $993$ link-instances,
used in full as a held-out set.

\begin{table*}[!tbp]
\centering
\caption{Per-parameter estimation statistics at the seven $(K,C)$ anchors at the named operating point $b=6$, jointly on Scene-Synth (in-scene, 2{,}400 test link-instances) and Scene-IITM (cross-scene, 993 link-instances, trained only on Scene-Synth). Each anchor occupies two rows; the second row reports the cross-scene evaluation. Detection is over the full test set; per-parameter errors are MAE and 50/90/99-th percentiles over detected link-instances.}
\label{tab:kc_anchors_full4_joint_b6}
\setlength{\tabcolsep}{2.5pt}
\renewcommand{\arraystretch}{1.05}
\resizebox{\textwidth}{!}{%
\begin{tabular}{r r l | r | r r r r | r r r r | r r r r | r r r r}
\toprule
\multirow{2}{*}{$K$} & \multirow{2}{*}{$C$} & \multirow{2}{*}{Dataset} &
\multirow{2}{*}{\makecell{Det\\(\%)}} &
\multicolumn{4}{c}{Delay $\varepsilon_\tau$ (ns)} &
\multicolumn{4}{c}{Doppler $\varepsilon_{fd}$ (Hz)} &
\multicolumn{4}{c}{Azimuth $\varepsilon_\phi$ ($^\circ$)} &
\multicolumn{4}{c}{Elevation $\varepsilon_\theta$ ($^\circ$)} \\
\cmidrule(lr){5-8}\cmidrule(lr){9-12}\cmidrule(lr){13-16}\cmidrule(lr){17-20}
& & & & MAE & P50 & P90 & P99 & MAE & P50 & P90 & P99 & MAE & P50 & P90 & P99 & MAE & P50 & P90 & P99 \\
\midrule
\multirow{2}{*}{4}  & \multirow{2}{*}{16} & Synth & 95.83 & 0.55 & 0.50 & 1.00 & 1.37 & 29.12 & 29.05 & 51.47 & 60.62 & 1.18 & 0.86 & 2.49 & 5.29 & 1.18 & 0.95 & 2.42 & 4.71 \\
                    &                     & IITM  & 99.19 & 0.56 & 0.47 & 0.98 & 1.38 & 29.41 & 29.60 & 52.96 & 60.30 & 1.05 & 0.84 & 2.16 & 3.73 & 1.21 & 0.96 & 2.56 & 4.25 \\
\midrule
\multirow{2}{*}{4}  & \multirow{2}{*}{32} & Synth & 96.33 & 0.54 & 0.51 & 0.98 & 1.48 & 28.47 & 28.32 & 50.70 & 58.97 & 0.90 & 0.65 & 1.92 & 5.21 & 0.77 & 0.61 & 1.64 & 2.97 \\
                    &                     & IITM  & 98.79 & 0.60 & 0.52 & 0.95 & 1.48 & 28.99 & 28.85 & 51.77 & 58.76 & 0.73 & 0.60 & 1.51 & 2.63 & 0.74 & 0.61 & 1.50 & 2.58 \\
\midrule
\multirow{2}{*}{8}  & \multirow{2}{*}{32} & Synth & 97.75 & 0.61 & 0.49 & 0.96 & 1.73 & 28.65 & 28.29 & 51.08 & 59.48 & 1.26 & 1.01 & 2.54 & 5.66 & 1.08 & 0.80 & 2.28 & 5.29 \\
                    &                     & IITM  & 99.09 & 0.54 & 0.48 & 0.92 & 1.68 & 28.94 & 28.77 & 52.70 & 58.07 & 1.08 & 0.85 & 2.10 & 4.25 & 0.97 & 0.77 & 2.05 & 3.52 \\
\midrule
\multirow{2}{*}{8}  & \multirow{2}{*}{64} & Synth & 98.17 & 0.57 & 0.49 & 0.99 & 1.36 & 28.54 & 28.22 & 50.88 & 58.20 & 1.05 & 0.77 & 2.24 & 5.15 & 0.84 & 0.68 & 1.73 & 3.20 \\
                    &                     & IITM  & 99.50 & 0.53 & 0.52 & 0.96 & 1.22 & 29.14 & 29.45 & 52.14 & 57.46 & 0.90 & 0.71 & 1.85 & 3.71 & 0.75 & 0.63 & 1.55 & 2.84 \\
\midrule
\multirow{2}{*}{16} & \multirow{2}{*}{32} & Synth & 98.29 & 0.54 & 0.50 & 0.95 & 1.21 & 28.55 & 28.38 & 50.98 & 58.93 & 1.25 & 0.95 & 2.60 & 5.90 & 1.19 & 0.88 & 2.56 & 5.52 \\
                    &                     & IITM  & 99.50 & 0.55 & 0.52 & 0.94 & 1.13 & 29.06 & 28.96 & 52.38 & 58.20 & 1.15 & 0.88 & 2.42 & 5.00 & 0.98 & 0.76 & 2.00 & 4.41 \\
\midrule
\multirow{2}{*}{16} & \multirow{2}{*}{64} & Synth & 98.88 & 0.53 & 0.51 & 0.93 & 1.28 & 28.56 & 28.46 & 50.79 & 58.74 & 1.05 & 0.82 & 2.15 & 4.78 & 0.77 & 0.59 & 1.61 & 3.31 \\
                    &                     & IITM  & 99.50 & 0.52 & 0.50 & 0.93 & 1.19 & 28.76 & 29.09 & 51.47 & 57.18 & 0.87 & 0.73 & 1.76 & 2.97 & 0.72 & 0.59 & 1.46 & 2.64 \\
\midrule
\multirow{2}{*}{64} & \multirow{2}{*}{32} & Synth & 99.21 & 0.52 & 0.50 & 0.92 & 1.16 & 28.54 & 28.59 & 51.16 & 57.90 & 1.17 & 0.94 & 2.33 & 5.07 & 0.92 & 0.73 & 1.88 & 4.40 \\
                    &                     & IITM  & 99.70 & 0.49 & 0.47 & 0.92 & 1.10 & 28.87 & 28.93 & 51.96 & 58.69 & 0.99 & 0.84 & 2.02 & 3.14 & 0.84 & 0.69 & 1.71 & 3.15 \\
\bottomrule
\end{tabular}%
}
\end{table*}

Table~\ref{tab:kc_anchors_full4_joint_b6} reports the seven
shortlist anchors at $b=6$ jointly on Scene-Synth (in-scene)
and Scene-IITM (cross-scene). Three observations emerge.

\emph{(i) Detection is preserved or improves on the unseen
scene.} At each named anchor, the Scene-IITM detection rate
is strictly greater than the corresponding Scene-Synth
detection rate, by $2.46$~pp (UC: $96.33 \rightarrow 98.79\%$),
$1.33$~pp (STD: $98.17 \rightarrow 99.50\%$), and $0.62$~pp
(HF: $98.88 \rightarrow 99.50\%$). The codec is not specialized
to Scene-Synth's clutter distribution.

\emph{(ii) Angular accuracy improves on Scene-IITM.} The
Scene-IITM azimuth and elevation MAEs are lower than the
corresponding Scene-Synth values at every named anchor. At STD,
azimuth MAE drops from $1.05^\circ$ to $0.90^\circ$ and
elevation MAE from $0.84^\circ$ to $0.75^\circ$. The tails
contract more strongly: at HF, the Scene-Synth azimuth $P99$
is $4.78^\circ$, while the Scene-IITM azimuth $P99$ is
$2.97^\circ$; the corresponding elevation $P99$ values are
$3.31^\circ$ and $2.64^\circ$. The plausible cause is that the
open-area campus layout exposes a sparser, less aggressive
interfering-scatterer field than the dense urban Scene-Synth;
the candidate-latent representation, having been trained to
encode target-specific rather than scene-specific structure,
benefits from the cleaner background at inference time.

\emph{(iii) Delay and Doppler statistics are scene-invariant.}
For every shortlist anchor, the cross-scene delay
MAE/$P90$/$P99$ values differ from the in-scene values by less
than $0.15$~ns, and the corresponding Doppler differences are
within $2$~Hz. These channels are limited by the OFDM
resolution rather than by clutter geometry, so they are
expected to be portable by construction; the data confirms it.

The combined effect is that the headline compression claim of
Sec.~\ref{sec:headline} extends to cross-scene operation
unchanged: at $107$/$404$/$806$~B per sample, the UC/STD/HF
anchors reach $98.79$/$99.50$/$99.50\%$ detection on Scene-IITM
with angular MAEs at or below $1.2^\circ$.

\subsection{Component ablations}
\label{sec:ablation_studies}

\begin{table*}[!tbp]
\centering
\caption{Component ablation study at the reference configuration $(K,C)=(256,8)$ with FP32 latents on Scene-Synth. Each row modifies a single component while keeping all others fixed. The ablated components (encoders, set transformer, reranker) operate on a per-token basis independent of $K$ and $C$, so the architectural conclusions drawn here carry over to all three named operating points (UC, STD, HF) of Table~\ref{tab:baseline_static_removal}. Detection rate is over the full test set ($2{,}400$ link-instances); errors are MAE and P50/P90 over detected link-instances.}
\label{tab:ablation_detected}
\setlength{\tabcolsep}{3pt}
\renewcommand{\arraystretch}{1.1}
\resizebox{\textwidth}{!}{%
\begin{tabular}{l l c ccc ccc ccc ccc}
\toprule
\multirow{2}{*}{Group} & \multirow{2}{*}{Variant} &
\multirow{2}{*}{Det.\ (\%)} &
\multicolumn{3}{c}{Delay $\varepsilon_\tau$ (ns)} &
\multicolumn{3}{c}{Doppler $\varepsilon_{fd}$ (Hz)} &
\multicolumn{3}{c}{Azimuth $\varepsilon_\phi$ ($^\circ$)} &
\multicolumn{3}{c}{Elevation $\varepsilon_\theta$ ($^\circ$)} \\
\cmidrule(lr){4-6}\cmidrule(lr){7-9}\cmidrule(lr){10-12}\cmidrule(lr){13-15}
& & & MAE & P50 & P90 & MAE & P50 & P90 & MAE & P50 & P90 & MAE & P50 & P90 \\
\midrule
Reference         & Full architecture with DD-context                              & 97.62 & 0.539 & 0.499 & 0.924 & 28.39 & 28.30 & 50.43 & 1.68 & 1.38 & 3.41 & 1.73 & 1.28 & 3.92 \\
\midrule
Reranking         & without reranker                                    & 75.42 & 2.843 & 1.175 & 4.925 & 32.78 & 29.78 & 57.21 & 1.47 & 1.12 & 3.03 & 1.27 & 0.91 & 2.92 \\
\midrule
\multirow{4}{*}{\makecell[l]{Token\\encoders}}
 & without angular encoder $g_{\mathrm{AE}}$                            & 15.54 & 1.178 & 0.558 & 1.086 & 30.28 & 30.41 & 52.77 & 4.62 & 4.27 & 8.60 & 4.41 & 3.99 & 8.82 \\
 & without position encoder $g_{\mathrm{pos}}$                          & 97.88 & 0.693 & 0.500 & 0.920 & 28.85 & 28.75 & 51.21 & 1.18 & 0.79 & 2.65 & 1.00 & 0.75 & 2.16 \\
 & without score encoder $g_{\mathrm{sc}}$                              & 97.75 & 0.578 & 0.472 & 1.015 & 27.86 & 28.03 & 49.36 & 1.88 & 1.32 & 4.31 & 1.45 & 1.15 & 3.00 \\
 & without DD-context encoder $g_{\mathrm{DD}}$                         & 98.83 & 0.565 & 0.511 & 0.933 & 28.27 & 28.29 & 50.26 & 1.14 & 0.81 & 2.56 & 1.05 & 0.78 & 2.23 \\
\midrule
\multirow{2}{*}{\makecell[l]{SF-side\\modules}}
 & without Set Transformer ($0$ layers)                                  & 97.38 & 0.776 & 0.584 & 1.316 & 28.39 & 28.35 & 50.98 & 1.52 & 1.15 & 3.32 & 1.47 & 1.09 & 3.14 \\
 & Set Transformer, $2$ layers                                           & 97.83 & 0.701 & 0.501 & 0.948 & 28.37 & 28.43 & 51.26 & 1.60 & 1.29 & 3.28 & 1.53 & 1.22 & 3.15 \\
\bottomrule
\end{tabular}%
}
\end{table*}
To attribute the observed gains to specific architectural
choices, Table~\ref{tab:ablation_detected} reports a
single-component ablation study at the reference configuration
$(K, C){=}(256, 8)$ with FP32 latents. The seven ablation
variants each modify exactly one element of the pipeline (a
token encoder, the Set Transformer, the reranker, or the
DD-context encoder) while keeping all other components at the
proposed setting. Because every ablated component is per-token
or per-set rather than $(K, C)$-capacity-dependent, the
conclusions drawn here transfer directly to the UC, STD, and
HF anchors of Table~\ref{tab:baseline_static_removal}.

\emph{Reranker is essential.} Removing the reranker (so that
the SF aggregates candidates with uniform weights) drops
detection from $97.62\%$ to $75.42\%$ and inflates delay MAE
from $0.54$~ns to $2.84$~ns. The proposal stage still returns
the correct Top-$K$ candidate set, but the SF can no longer
extract the right hypothesis from it. This is large delta in the ablation and is consistent with the
intended coarse-to-fine design: high proposal coverage is
necessary but candidate-level reranking is what converts that
coverage into accurate final estimation.

\emph{Angular encoder $g_{\mathrm{AE}}$ dominates among the
token encoders.} Removing $g_{\mathrm{AE}}$ collapses detection
to $15.54\%$ and more than doubles the azimuth and elevation MAEs
(to $4.62^\circ$ and $4.41^\circ$, i.e. $2.75\times$ and $2.55\times$
the reference values). Without it, the candidate
token carries almost no angular information and the SF has no
basis on which to estimate $\phi$/$\theta$. Removing the
position encoder $g_{\mathrm{pos}}$ or the score encoder
$g_{\mathrm{sc}}$ leaves detection within $0.3$~pp of the
reference; their contribution is concentrated in sub-bin delay
refinement (position) and reranking calibration (score).

\emph{Cross-candidate reasoning helps at sub-bin scale.}
Removing the single Set Transformer layer (so candidates are
fed directly to the reranker without cross-candidate attention)
preserves detection at $97.38\%$ but degrades delay MAE to
$0.78$~ns; angular MAEs and $P90$ tails vary only within seed
noise (e.g., $P90$ azimuth shifts from $3.41^\circ$ to
$3.32^\circ$). Doubling to two layers does not move any metric
beyond seed variability, so a single layer is retained.

\emph{Removing the DD-context encoder improves detection.} At
the reference $(K, C){=}(256, 8)$ point, the no-DD-context
variant reaches $98.83\%$ detection versus $97.62\%$ for the
full architecture. The DD-context encoder appears to add
redundant capacity that the model uses inefficiently in
single-target operation; the no-DD-context architecture is
therefore adopted as the backbone for the $(K, C)$ sweep of
Sec.~\ref{sec:design_studies} and for the UC, STD, HF
anchors.

The ablation table therefore supports the proposed pipeline as:
power$+$peakedness DD descriptor, three-encoder candidate
token ($g_{\mathrm{AE}}$, $g_{\mathrm{pos}}$,
$g_{\mathrm{sc}}$), single-layer Set Transformer, candidate
reranker, and bounded sub-bin offset head, with the
DD-context encoder removed for single-target operation.

\subsection{Comparison with related ISAC estimators in the literature.}
We summarise reported estimation accuracy from ISAC works to position the proposed pipeline qualitatively.
Quantitative cross-paper comparison of delay and Doppler errors is
fundamentally constrained by geometry: in bistatic operation, delay
maps to an ellipsoidal locus rather than a range and Doppler depends on
the bistatic angle, so monostatic range/velocity error figures from the
literature are not directly comparable to bistatic delay/Doppler errors
without per-sample geometric reduction. We therefore restrict
cross-paper comparison to angular accuracy and to qualitative scope of
the estimation problem.

Among learning-based ISAC estimators, Hu \emph{et al.}~\cite{Hu2024TwoStageISACReceiver}
(STransformer + MUSIC, SIMO-OFDM, $N_r{=}8$ ULA, 64 subcarriers,
\emph{static} target) report an AoA error of approximately $0.1^\circ$
at SNR~$\geq 6$~dB but estimate only AoA and time delay; neither
Doppler nor elevation is addressed. ISAC-NET~\cite{Jiang2024ISACNET}
(model-driven unfolding, single-antenna passive sensing) improves
communication BER and range/velocity RMSE over a 2D-DFT baseline but
estimates no angular parameter. Among recent classical 5G-NR-aligned
estimators, the joint 3D-DFT method of Xiao
\emph{et al.}~\cite{Xiao2024JointARV} ($16{\times}16$ ULA, 512 subcarriers,
CPI~$L{=}256$) and the auto-paired super-resolution 3DJE of Hu
\emph{et al.}~\cite{HuZelin2024} estimate at most three parameters
(range/velocity/azimuth) over ULA configurations and do not address
elevation.

Among model-based clutter-aware ISAC frameworks, the closest comparable
work is Luo \emph{et al.}~\cite{Luo2024ClutterEnvironment}, which performs joint
angle/distance/velocity estimation under static ground clutter via
\emph{explicit} mean-phasor cancellation across OFDM symbols followed by
angle--Doppler spectrum estimation and a MUSIC-based range/velocity
stage. That work reports angular RMSE down to $\sim 0.01^\circ$ under monostatic 128-element TX/RX ULAs at 220 GHz, but does not estimate elevation, assumes explicit
static-clutter filtering as a prerequisite, and is evaluated on a
synthetic clutter model rather than ray-traced multipath. The proposed
pipeline differs from~\cite{Luo2024ClutterEnvironment} by handling clutter
\emph{implicitly} through the learned encoder operating directly on the
raw DDAE magnitude features, by adding elevation
as a fourth jointly estimated parameter via UPA aperture, by operating
in a bistatic ray-traced scene with both static and dynamic clutter,
and by exposing the estimation problem to a finite SE--SF feedback
budget that none of the above works addresses.

As anticipated in Sec.~\ref{sec:introduction}-A, no previously published ISAC 
estimator -- whether learning-based or model-based -- jointly estimates
delay, Doppler, azimuth, \emph{and} elevation over a UPA in a multipath
bistatic scene with static and dynamic clutter under a finite per-link
feedback budget, which is the operating regime considered here. At the
STD anchor with $b{=}6$, the proposed pipeline achieves azimuth and
elevation MAEs of $1.05^\circ$ and $0.84^\circ$ respectively, which are
in the same order as the sub-degree angular accuracies reported
in~\cite{Hu2024TwoStageISACReceiver,Xiao2024JointARV,HuZelin2024,Luo2024ClutterEnvironment} despite the more demanding
joint four-parameter, UPA, ray-traced bistatic, clutter setting and the
finite feedback-budget constraint.

\subsection{Summary and limitations}
\label{sec:design_takeaways}

Three takeaways summarize the results of this section. First,
the proposed candidate-latent codec reduces the per-link
SE--SF feedback to only 107--806 bytes per CPI across the
UC, STD, and HF anchors, while preserving 96.3--98.9\%
detection rate on Scene-Synth and 98.8--99.5\% on Scene-IITM,
with delay MAE below 0.6 ns, Doppler MAE near 28.5 Hz,
and angular MAEs below 1.2$^\circ$ across the three named
anchors. For context, direct forwarding would require
107.35 MB per CPI for raw 16-bit-I/16-bit-Q CFR, 39.32 MB
for FP32 DDAE magnitude, and 9.83 MB even for 8-bit DDAE
magnitude. Second, the codec admits substantial
post-training quantization: $b=6$ is essentially free across
all evaluated $(K, C)$ anchors, and $b=4$ remains operational. Third, the trained representation is portable: it transfers to an
independent campus-scale scene without retraining and in fact
achieves tighter angle errors on the unseen scene than on the
training distribution.

Several limitations bound the scope of these conclusions and
motivate future work. The codec is evaluated only in the
single-target regime; multi-target settings introduce
candidate-association ambiguity and require an extension of
the reranker beyond the current single-target attention head.
The quantization study is post-training; quantization-aware
training, adaptive bit allocation across the $(r, k)$ axes,
and entropy coding of the latent and candidate-index streams
are not explored. Cross-scene generalization is demonstrated
on a single OOD scene; broader generalization to dense urban,
indoor, or larger-scale environments is not assessed. Finally,
the SE--SF interface is studied in isolation: end-to-end
multistatic fusion, fronthaul latency, scheduling, and joint
communication--sensing resource allocation are outside the
scope of this paper. These directions are discussed further in
Sec.~\ref{sec:conclusion}.

\section{Conclusion}
\label{sec:conclusion}

We address the SE--SF feedback-interface problem in distributed OFDM-ISAC sensing by introducing a coarse-to-fine candidate-latent codec. The SE proposes $K$ delay--Doppler candidates and encodes each into a $C$-dimensional latent token; the quantized $K\times C$ latent matrix, the candidate indices, and one global scale together constitute the per-link feedback. The SF dequantizes, applies cross-candidate refinement, reranks, and produces final estimates of delay, Doppler, azimuth, and elevation.

Across the three named operating points, the trained codec reduces the
per-link SE--SF feedback to only 107--806 bytes per CPI, while maintaining
96.3--98.9\% detection on the Scene-Synth held-out test set. For context,
direct forwarding would require 107.35 MB per CPI for raw
16-bit-I/16-bit-Q CFR, 39.32 MB for FP32 DDAE magnitude, and 9.83 MB
even for 8-bit DDAE magnitude. The latent representation tolerates $6$-bit post-training quantization at essentially no detection cost, and $4$-bit at moderate angular-tail cost.  Cross-scene evaluation on Scene-IITM, an independent campus-scale environment with no scene-specific fine-tuning, preserves detection ($98.8$--$99.5\%$ across the three anchors) and tightens the angular error distributions, indicating that the learned representation encodes target-specific rather than scene-specific structure. Component ablations attribute the gains to candidate-level reranking and the angular token encoder, with smaller but consistent contributions from position-encoded sub-bin refinement and cross-candidate reasoning via a single-layer Set Transformer.

The framework as presented is restricted to single-target
estimation, post-training quantization with one global scale,
and a single in-scene/out-of-scene pair. Practically important
extensions include multi-target association at the sensing
fusion side, quantization-aware training and entropy coding of
the candidate-latent stream, learned bit allocation across the
candidate-index, latent, and side-information channels, and
full multistatic fusion under fronthaul-latency and joint
communication--sensing resource constraints. Embedding the
proposed codec into such an end-to-end distributed
integrated sensing and communication system is the natural
next step toward deployment.

\begingroup
\sloppy
\bibliographystyle{IEEEtran}
\bibliography{references}
\endgroup

\begin{IEEEbiography}[{\includegraphics[width=1in,height=1.25in,clip,keepaspectratio]{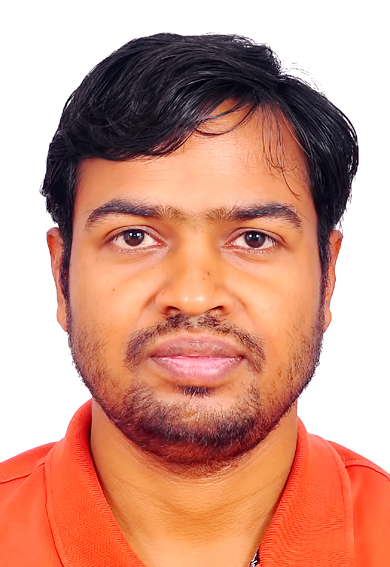}}]{Shiv Shankar}\hspace{0.4em}received the M.Tech.\ degree in electronics and communication engineering from IIT (ISM) Dhanbad. He is currently pursuing the Ph.D.\ degree with IIT Madras, Chennai, India, and is a Principal Research Engineer with the CEWiT, IIT Madras. His research interests include AI/ML for wireless systems, integrated sensing and communication (ISAC), and digital-twin-enabled network intelligence. He has also contributed to 3GPP work items on AI/ML for positioning and CSI enhancement.
\end{IEEEbiography}

\begin{IEEEbiography}[{\includegraphics[width=1in,height=1.25in,clip,keepaspectratio]{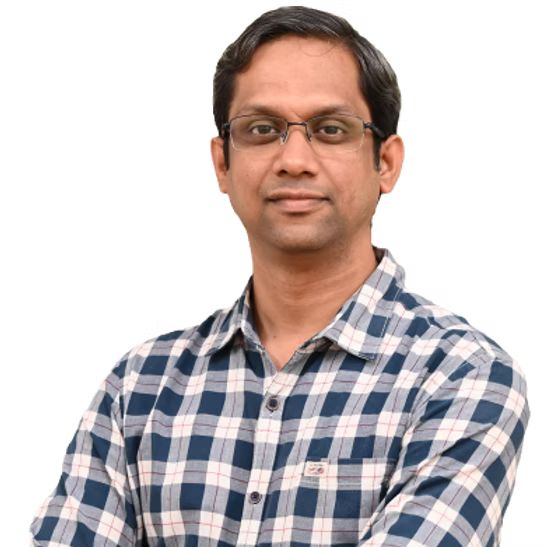}}]{Radha Krishna Ganti}\hspace{0.4em}received the B.Tech.\ and M.Tech.\ degrees in electrical engineering from IIT Madras, Chennai, India, and the M.S.\ degree in applied mathematics and the Ph.D.\ degree in electrical engineering from the University of Notre Dame, USA, in 2009. He is currently a Professor with IIT Madras. His research interests include wireless networks, AI/ML-based receivers, ISAC, and next-generation wireless systems. He was a recipient of the Rashtriya Vigyan Yuva--Shanti Swarup Bhatnagar Award in 2024.
\end{IEEEbiography}

\begin{IEEEbiography}[{\includegraphics[width=1in,height=1.25in,clip,keepaspectratio]{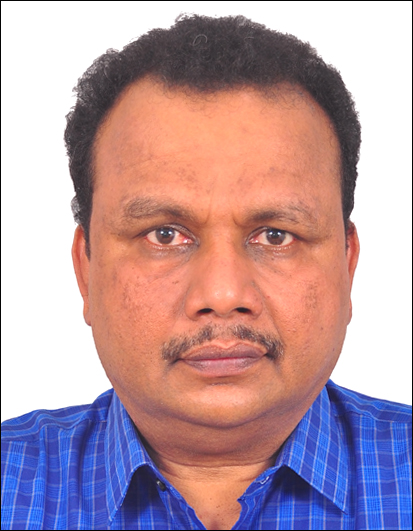}}]{J Klutto Milleth}\hspace{0.4em}received the Ph.D.\ degree in space-time wireless communication from IIT Madras, Chennai, India, in 2004. He is currently the Chief Technologist with the CEWiT, IIT Madras, and an Adjunct Faculty Member with the Department of Electrical Engineering, IIT Madras. His work spans 5G and beyond wireless technologies, standardization, and early 6G research. He has contributed to major standards including 3GPP LTE/NR, IEEE 802.16m, and TSDSI 5Gi.
\end{IEEEbiography}

\end{document}